\documentclass[fleqn,usenatbib]{mnras}

\usepackage[T1]{fontenc}

\DeclareRobustCommand{\VAN}[3]{#2}
\let\VANthebibliography\thebibliography
\def\thebibliography{\DeclareRobustCommand{\VAN}[3]{##3}\VANthebibliography}

\usepackage{graphicx}	
\usepackage{amsmath}	
\usepackage{amssymb}	

\graphicspath{{./}{figures/}}
\usepackage{color,subfigure,lineno} 

\usepackage{eso-pic}

\usepackage{newtxtext,newtxmath}

\usepackage{float}
\usepackage{ulem}
\usepackage{soul}

\title[Quasar Accretion Disc Variability]{Temperature Fluctuations in Quasar Accretion Discs from Spectroscopic Monitoring Data}

\author[Stone \& Shen]{Zachary Stone$^{1,2}$\thanks{E-mail: stone28@illinois.edu (ZS)}
and Yue Shen$^{1,3}$
\\
$^{1}$Department of Astronomy, University of Illinois at Urbana-Champaign, Urbana, IL 61801, USA\\
$^{2}$Center for AstroPhysical Surveys, National Center for Supercomputing Applications, University of Illinois at Urbana-Champaign, Urbana, IL 61801, USA\\
$^{3}$National Center for Supercomputing Applications, University of Illinois at Urbana-Champaign, Urbana, IL 61801, USA}

\date{Accepted XXX. Received YYY; in original form ZZZ}

\pubyear{2023}

\begin{document}
\label{firstpage}
\pagerange{\pageref{firstpage}--\pageref{lastpage}}
\maketitle

\begin{abstract}
Neustadt \& Kochanek (2022, hereafter NK22) proposed a new method to reconstruct the temperature perturbation map (as functions of time and disc radius) of AGN accretion discs using multi-wavelength photometric light curves. We apply their technique to 100 quasars at $z=0.5-2$ from the Sloan Digital Sky Survey Reverberation Mapping project, using multi-epoch spectroscopy that covers rest-frame UV-optical continuum emission from the quasar and probes days to months timescales. Consistent with NK22 for low-redshift AGNs, we find that the dominant pattern of disc temperature perturbations is either slow inward/outward moving waves with typical amplitudes $\delta T/T_0\sim 10\%$ traveling at $\sim 0.01-0.1c$, with a typical radial frequency of $\sim$ 0.5 dex in $\log R$, or incoherent perturbations. In nearly none of the cases do we find clear evidence for coherent, fast outgoing temperature perturbations at the speed of light, reminiscent of the lamppost model; but such lamppost signals may be present in some quasars for limited periods of the monitoring data. Using simulated data, we demonstrate that high-fidelity temperature perturbation maps can be recovered with high-quality monitoring spectroscopy, with limited impact from seasonal gaps in the data. On the other hand, reasonable temperature perturbation maps can be reconstructed with high-cadence photometric light curves from the Vera C. Rubin Observatory Legacy Survey of Space and Time. Our findings, together with NK22, suggest that internal disc processes are the main driver for temperature fluctuations in AGN accretion discs over days to months timescales. 
\end{abstract}

\begin{keywords}
accretion discs -- galaxies: active
\end{keywords}



\section{Introduction} \label{sec:intro}

It is well accepted that the typical stochastic variability observed in the UV-optical continuum emission of active galactic nuclei (AGN) stems from thermal emission in the accretion disc immediately surrounding its supermassive black hole (SMBH). This emission occurs at all radii within the accretion disc, with blue, hot emission on average originating at smaller radii and red, cooler emission coming from larger radii. However, the origin of accretion disc variability is still uncertain.  

Empirically, optical AGN variability is characterized well by a number of statistical models, including the Damped Random Walk model \citep{2010ApJ...721.1014M, 2021ApJ...907...96S, stone_optical_2022}, or the higher-order CARMA models \citep{kelly_flexible_2014, Simm_etal_2016, moreno_stochastic_2019, yu_examining_2022}. In addition, there are a number of physical models to explain AGN time-series data, such as the lamppost model \citep{friedjung_accretion_1985, sergeev_lag-luminosity_2005} and the CHAR model \citep{sun_corona-heated_2020}. In order for observations to match the thermal emission from an accretion disc, it is often assumed that there is a hot, X-ray-emitting, variable corona near the center along the axis of rotation \citep{galeev_structured_1979, haardt_two-phase_1991, chakrabarti_spectral_1995}. This X-ray corona pairs with the lamppost model, stating that hot X-rays emitted close to the SMBH are reprocessed by the rest of the disc \citep{zdziarski_electron-positron_1990, haardt_two-phase_1991, frank_accretion_2002}. Temperature fluctuations from the corona region that propagate through the disc at the speed of light are then observed as flux variability in time-series observations. For these fluctuations to be propagating outward, emission at smaller radii would lead emission at larger radii. Thus, X-ray emission would lead UV emission, which would lead optical emission, etc.

The lamppost model is utilized to map the geometry of the AGN through time lags with the reverberation mapping method \citep{blandford_reverberation_1982, peterson_reverberation_1993,Cackett_etal_2021}. Many studies have correlated AGN optical light curves in different bands to measure this time lag between propagating fluctuations within the accretion disc, also assuming a thin disc model \citep[][hereafter SSD]{lynden-bell_galactic_1969, pringle_accretion_1972, shakura_black_1973, novikov_astrophysics_1973}. As the observed wavelength of the light curve increases, the time lag would as well \citep[e.g.,][]{collier_new_1999}. Many studies indeed find UV-leading-optical behavior in the continuum light curves of local AGNs \citep{cackett_testing_2007, cackett_accretion_2018, edelson_space_2015, edelson_lessigreaterswiftlessigreatermonitoring_2017, fausnaugh_space_2016, sergeev_lag-luminosity_2005, mchardy_swift_2014}. 

However, there are other studies that present evidence to counter the lamppost model. Some studies find that X-ray emission leads optical emission \citep{shappee_man_2014, arevalo_correlation_2009, breedt_twelve_2010, troyer_correlated_2016}, while others find no correlation \citep{maoz_x-ray_2002}, or an anti-correlation \citep{shemmer_complex_2003, marshall_correlated_2008}. In most cases this X-ray/optical correlation is weak and produces time lags that cannot be fully explained by the simplest lamppost models \citep{edelson_lessigreaterswiftlessigreatermonitoring_2017, edelson_first_2019}. Many studies also find that the size of the accretion disc is too large compared to the predicted size in the SSD model by a factor of $\sim 3$ \citep{morgan_quasar_2010, edelson_space_2015, edelson_lessigreaterswiftlessigreatermonitoring_2017, fausnaugh_space_2016, cackett_accretion_2018}, but in some cases the inferred disc sizes from continuum reverberation mapping could be biased high by diffuse emission from the broad-line region \citep[e.g.,][]{chelouche_direct_2019, Guo_etal_2022}. Recently, there have been a number of suggested solutions to this issue, including bias in luminosity from nuclear extinction \citep{gaskell_estimating_2023}, underestimated black hole mass from unknown structure in the broad-line region \citep{pozo_nunez_optical_2019}, non-blackbody emission from the disc due to scattering in its atmosphere \citep{hall_non-blackbody_2018}, and disc inhomogeneity \citep{dexter_quasar_2010}.

There is evidence that the observed variability is a combination of both outward-moving lamppost-like signals and inward-moving signals as well \citep{hernandezsantisteban_intensive_2020}. \cite{dexter_quasar_2010} suggest that an inhomogeneous disc could create thermal fluctuations that propagate inward. This model accounts for observed optical variability, as well as the discrepancy in the disc size between the SSD prediction and observations. Such an inhomogeneous disc could be formed through various instabilities (e.g., MRI, thermal, viscous). \cite{lyubarskii_flicker_1997} suggest that changes in viscosity throughout the disc could cause flux variations at lower radii to follow those at larger radii. \cite{hernandezsantisteban_intensive_2020} find evidence for slow-moving inward-propagating perturbations over longer timescales, in addition to the lamppost-like signals observed on much shorter timescales.

\citet[][hearafter NK22]{Neustadt_2022} have proposed a method to reconstruct temperature perturbation maps of AGN accretion discs directly from time-series observations. Their method relies on a minimal number of assumptions, only assuming axisymmetric emission from a Shakura-Sunayev thin disc and linear temperature perturbations within the disc. NK22 utilize high-cadence multi-band light curves for 7 nearby AGN. They find evidence for low-amplitude lamppost-like temperature fluctuations in most of their sample, consistent with the blue-leading-red behavior seen in their light curves. More interestingly, NK22 find that the dominant mode of temperature fluctuations is slow, ingoing and outgoing waves with a larger amplitude than the lamppost-like signal. These slow wave perturbations decrease in velocity when radius decreases, reaching $\sim 0.01c$ close to the SMBH. 


Most prior studies of emission from AGN accretion discs assume that the AGN is at a sufficient distance where it can be considered a point source. However, this fails to account for the light travel time differences between different parts of the disc, if viewed from a non-face-on configuration.  Emission from the far side of the disc will be observed together with emission from the near side of the disc with a light travel time delay. This delay causes emission from the disc to be ``smeared out" over a range of observed times. This smearing also depends on the parameters of the disc (i.e. the extent), and hence should be included in model analysis of time-series and spectroscopic data. NK22 properly takes in account this ``smearing effect'' due to light-travel time in their method.

To further understand the propagation of temperature fluctuations in AGN accretion discs, here we apply the NK22 method to a sample of distant quasars from the Sloan Digital Sky Survey Reverberation Mapping project \citep[e.g.,][]{Shen_etal_2015a}, with multi-epoch spectroscopy that covers rest-frame UV-optical emission. While our spectroscopic light curves do not have the same high cadence as the photometric light curves used in NK22, the coverage in wavelength is continuous, and with the SDSS-RM sample we are able to extend to a much larger sample of quasars beyond the nearby Universe.


In Section \ref{sec:methods}, we describe the approach in NK22. We retain the essential NK22 formalism for notation and reference purposes, and we refer the reader to NK22 for full details of this approach. In Section \ref{sec:tests}, we perform a suite of tests on the algorithm with simulated multi-epoch spectra. In Section \ref{sec:real_data}, we describe the SDSS-RM sample, and the reconstructed temperature perturbation maps using monitoring spectroscopy. We conclude in \S\ref{sec:conclusion}.

\section{The NK22 Method} \label{sec:methods}


Here, we describe the method utilized in NK22 to reconstruct the temperature perturbation map in the two-dimensional radius-time ($R-t$) space, using input light curve data. While the formalism is from NK22, we repeat it below for better flow of the paper \footnote{We have corrected a few misprints in the original NK22 formalism (Eqns.~6 \& 7) and added more technical details in Appendix~\ref{sec:smearing_deriv}. The actual analyses in NK22 were using the correct formulae (J.~Neustadt, private communications).}. As stated in NK22, we assume a steady-state AGN accretion disc with a given a radial temperature profile $T(R)$ from SSD, and consider linear temperature perturbations. Axisymmetry is assumed for both the steady state and the perturbations. These temperature fluctuations are observed as variations of the emitted flux $F_\lambda (\lambda, t)$ on top of the steady-state flux distribution $F_{\lambda, SS}(\lambda, t)$. The SSD has a radial temperature profile of $T \propto R^{-3/4}$ and a flux spectrum $F_\nu \equiv dF / d\nu \propto \nu^{1/3}$. But it is straightforward to adopt a different steady-state temperature profile \citep[e.g.,][]{Weaver_Horne_2022}. 

An unperturbed disc has an inner radius $R_{in}$ at the Innermost Stable Circular Orbit (ISCO):
\begin{flalign}
    & R_{in} = \alpha R_g = \frac{\alpha G M_{BH}}{c^2} &
\end{flalign}
where $R_g$ is the gravitational radius and $\alpha$ is a parameter governing the location of the inner radius with respect to the SMBH. For a Schwarzschild BH, the ISCO occurs at $\alpha=6$ \citep{misner_gravitation_1973}, which we will assume going forward. The steady-state radial temperature profile for a standard thin disc as a function of the dimensionless radial coordinate $u \equiv R / R_{in}$ is: 
\begin{flalign}
    & T_0(u) = T_{in} u^{-3/4} (1 - u^{-1/2})^{1/4} \,\, \, {\rm where} & \tag{2a, 2b} \\
    & T_{in} = ( 1.54 \times 10^5 ) \lambda_{\rm Edd}^{1/4}  \left( \frac{10^9 M_{\odot}}{M_{BH}} \right)^{1/4}  \left( \frac{6}{\alpha} \right)^{3/4} {\rm K}& \nonumber
\end{flalign}
and $\lambda_{\rm Edd}\equiv L_{\rm bol}/L_{\rm Edd}$ is the Eddington ratio. 

The emitted flux of the accretion disc at a given wavelength $\lambda$ is given by integrating the Planck distribution over the entire disc. In terms of $u$:
\begin{flalign}
    F_\lambda = & \, F_{\lambda, 0}(\lambda) \int_1^\infty \frac{u \ du}{e^x - 1} & \tag{3a}\label{eqn:flam}\\
    {\rm where} & \,\, F_{\lambda, 0}(\lambda) = \frac{4 \pi h c^2 \cos(i) R_{in}^2}{\lambda^5 D^2} & \tag{3b}\\
                & \,\, x = \frac{hc}{\lambda k_B T_0(u)} & \tag{3c}
\end{flalign}

where $D$ is the distance to the AGN, $i$ is the inclination of the disc with respect to our line of sight, and we assume a radiative efficiency of $\eta = 0.1$ in deriving the temperature profile.


Assuming linear temperature perturbations $\delta T(u,t)$, the resulting change in flux is
\begin{flalign}\label{eqn:delta_flam}
    & \delta F_\lambda(t) = F_{\lambda, 0}(\lambda) \int \frac{u \ du}{(e^x - 1)^2} \frac{x e^x}{T_0(u)} \ \delta T(u, t) & \tag{4}
\end{flalign}

The disc can then be discretized into a matrix $W$, gridded in radius $u$ and time $t$. Each individual element of this matrix will then be the integrand, at that time and position in the disc. However, this does not account for the time it takes for light to travel across the accretion disc, which depends on both its size and inclination with respect to the observer. Light emitted at a certain time in the rest-frame of the AGN will be smeared to influence the flux received in a range of observed times. Consequently, flux perturbations at a given radius $u$ in NK22's model parameter time $t_p$ will be smeared out over a certain range of data times $t_d$, governed by a smearing function $f(u, t_p, t_d)$. This can be multiplied by the existing $W$ matrix to account for the smearing effect in Eqn.~\ref{eqn:delta_flam}.


The smearing term can be constructed by first defining the characteristic time-scale of light travel across the disc:
\begin{flalign}
    & t_0 = \frac{R \sin(i)}{c} = \frac{u R_{in} \sin(i)}{c} & \tag{5}
\end{flalign}

Define the smearing functions from NK22:
\begin{flalign}
    & G_1( x, y ) = \frac{1}{\pi \Delta t} \left( \sqrt{t_0^2 - y^2} - \sqrt{t_0^2 - x^2} \right) & \tag{6a}\label{eqn:G_term1} \\
    & G_2(x, y) = \frac{1}{\pi}\left[ {\rm arcsin}\left( \frac{y}{t_0} \right) - {\rm arcsin}\left( \frac{ x }{t_0} \right) \right] & \tag{6b}\label{eqn:G_term2}
\end{flalign}

\noindent
producing
\begin{flalign}\label{eqn:smearing_tot}
    f(u, t_p, t_d) \;\; & \!\! = \ - G_1(t_1, t_2) + \frac{t_d - (t_p - \Delta t)}{\Delta t} \;  G_2(t_1, t_2) \tag{7} \\
                        & \!\! + \ G_1(t_3, t_4) + \frac{(t_p + \Delta t) - t_d}{\Delta t} \; G_2(t_3, t_4) \nonumber
\end{flalign}
 
where
\begin{flalign}\label{eqn:int_bounds}
    & t_1 = {\rm max}(t_p - \Delta t - t_d, -t_0) & \tag{8a} \\
    & t_2 = {\rm min}(t_p - t_d, +t_0) & \tag{8b} \\
    & t_3 = {\rm max}(t_p - t_d, -t_0) & \tag{8c} \\
    & t_4 = {\rm min}(t_p + \Delta t - t_d, +t_0) & \tag{8d} 
\end{flalign}

where $\Delta t$ is the $t_p$-grid spacing. An in-depth derivation of this smearing function is given in Appendix \ref{sec:smearing_deriv}.

The purpose of this model is to construct temperature profile maps $\delta T(u, t_p)$ from spectroscopic/photometric light curve data of flux perturbations $\delta F_{\lambda}(\lambda, t_d)$. Therefore, we need to minimize $\chi^2$ in order to fit the model to our input flux data. To ensure robust matrix inversion, linear regularization is used to reduce the likelihood of large jumps in temperature perturbations in both the radial and temporal dimensions. Adding this ``smoothing" effect also remedies for situations where the matrix inversion is ill-conditioned or undefined. NK22 define a smoothing parameter $\xi$, which applies to all smoothing matrices (one for each dimension and a linear one). The choice of $\xi$ is made on a case-by-case basis, but as stated in NK22, usually $\xi$ is chosen so that the $\chi^2$ per data point ($\chi^2 / N_d$) is $\sim 1$, where $N_d=N_{t_d}\times N_\nu$ is the total number of data points.

Using this smoothing parameter $\xi$ and the smoothing matrices defined in NK22 ($D_k$, $D_l$, $I$), the $\chi^2$ of this model can be minimized to yield:
\begin{flalign}\label{eqn:inversion}
    & \delta T = \left[ W_\sigma^T W_\sigma + \xi (I + D_k + D_l) \right]^{-1} W_\sigma^T \delta F_\sigma & \tag{9}
\end{flalign}
\noindent
where $W_\sigma = W/\sigma$, $\delta F_{\sigma} = \delta F_{\lambda} / \sigma$, and $\sigma$ is the error in the observed flux perturbations. 


This inversion process maps two dimensions given in the observed data $(\lambda, t_d)$ to two dimensions in the temperature profile map $(u, t_p)$. The sampling of both $\lambda$ and $t_d$ are determined by the observed spectroscopic/photometric light curves. The sampling of $u$ and $t_p$ are determined on a case-by-case basis, both of which determine the quality of the $\delta T(u, t_p)$ fit. In general, the $t_p$ array must span the observed times $t_d$ with $N_{t_p}$ data points and a grid spacing of $\Delta t$. NK22 state that the $t_p$ array should go from the minimum $t_d$ value to the maximum $t_d$ value. However, knowing that the smearing term involves times before the first $t_d$ value and after the last $t_d$ value, we opt to use a $t_p$ array that starts a few days before the first observed time and ends a few days after the last observed time. In addition, all data must be shifted to the rest-frame of the AGN, including the observed times $t_d$, wavelengths $\lambda$, and fluxes $\delta F_\lambda$.

\begin{figure}
    \centering
    \includegraphics[width=.5\textwidth]{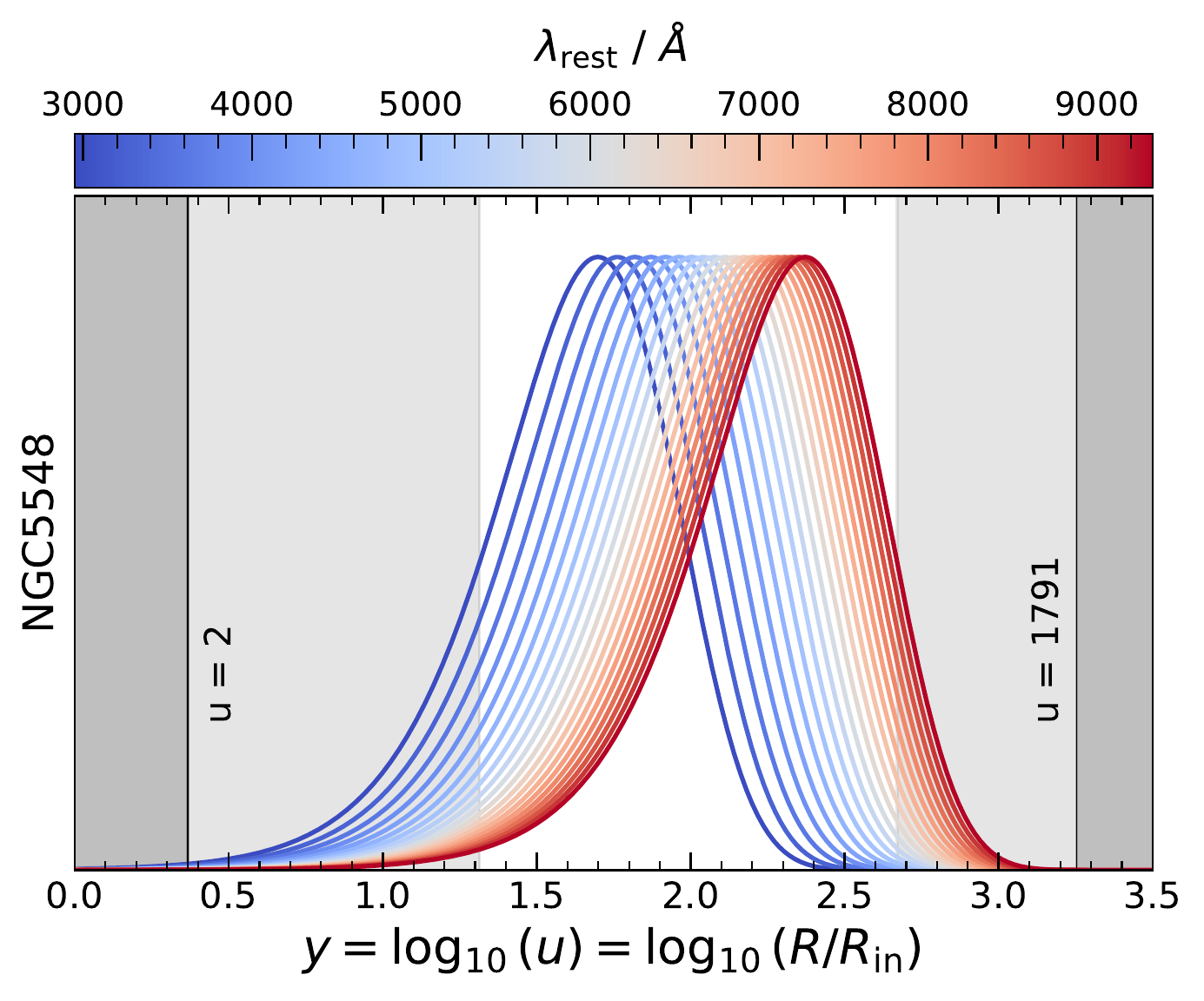}
    \caption{The filter kernels used for an AGN with the same accretion parameters as NGC5548 (described in \S \ref{sec:methods}), using Eqn.~\ref{eqn:delta_flam}, the parameters of the quasar, and the wavelength range described for our test cases in \S \ref{subsec:test_cases}. The range of radii probed are labeled with text and a dark line on either side of the kernels, with the rest of the parameter-space shaded in dark gray. The more restrictive radius range is shown, shading the rest of the parameter-space with a lighter gray.}
    \label{fig:kernels}
\end{figure}

To construct the grid in $u$-space, we undergo the same process of constructing filter kernels as in NK22 (Fig.~\ref{fig:kernels}). These filter kernels describe the weighted contribution of the temperature perturbation $\delta T$ as a function of radius on the observed flux perturbations $\delta F_\lambda$ at a given wavelength, obtained from the integrand in Eqn.~\ref{eqn:delta_flam}. This process helps to define the bounds of the integral used in Eqn.~\ref{eqn:delta_flam}, as to have the minimal amount of elements in the design matrix $W$. While NK22 uses the same range of radii for each object, we use different radii for each object, but keep the same resolution ($N_u$) and sample evenly in $\log_{10}(u) = y$. To obtain the range of radii explored by the data, we obtain the filter kernels for the shortest and longest rest-frame wavelengths probed by data for a given object. We designate the upper and lower bounds of the radii for each kernel at an arbitrary fraction of 0.01 of the kernel peak. The lower (upper) bound for the shortest-wavelength (longest-wavelength) kernel then defines the range of radii sampled by the data. For each object, we can define a more restrictive radius range, defined at half of the filter kernel maximum for the longest and shortest wavelengths. In terms of resolution, we follow NK22, and use $N_u = 50$ radial bins for each object, as the radial resolution has little effect on the quality of the output temperature profile map. We assume an inclination of $i = 30^o$ as well, as it also has little effect on the output.

An important caveat to this process is that the data we begin with are light curves in terms of the total flux $F_\lambda$, not the flux variations $\delta F_\lambda$. Subtracting the mean spectrum/light curve to obtain flux variations from this data manually can lead to a difference between the mean/steady-state spectrum in the model and the data, resulting in poor temperature map quality or an offset between the input and output light curves. This difference can stem from host-galaxy contamination, systematic offsets in the data, deviations from the thin disc model, etc. Therefore, as in NK22, we define a parameter for each light curve (i.e., at a specific wavelength) to account for this difference in offset between the model and the data. These offsets can be used to reconstruct light curves from the output temperature profile maps as well.


\section{Implementation of The NK22 Method} \label{sec:tests}

When creating input light curves from an input temperature profile map, we opt to use Eqn.~\ref{eqn:flam}, utilizing $T_0 + \delta T$ instead of $T_0$, because it allows the light curves to be created with a nonlinear dependence on the temperature perturbations. Therefore, results would show that this inversion process is robust with both linear and nonlinear trends, as it assumes that the light curves are linearly proportional to the temperature perturbations. The benefit of using these non-linearized light curves is discussed in NK22: the linearized light curves can produce unphysical negative temperatures and fluxes if the amplitude of perturbations becomes too large. 

In order to compare the input and output light curves and spectra, a normalization is needed. NK22 overcome this issue by normalizing the input and output light curves between $-1$ and 1. However, if the input/output contain too much noise, this method of normalization can be offset by a certain amount. We found it best to use the fitted offset parameters (one for each wavelength) and minimize $\chi^2$ to obtain a multiplicative factor to fit the output data to the input data. In addition, normalization is needed to compare the input and output temperature map profiles. As in NK22, we normalize the temperature maps by the 99$^{\rm th}$ percentile of $|\delta T / T|$ (labeled as the scale in each of the figures). 

We verify our implementation of the NK22 method by testing on the photometric light curves for nearby AGNs used in NK22, and reproduce the exact results. Next, we test the performance of the NK22 approach, using simulated temperature perturbation maps and the corresponding spectroscopic light curves. 


\subsection{Idealized Test Cases} \label{subsec:test_cases}



\begin{figure*}
    \centering
    \includegraphics[width=\textwidth]{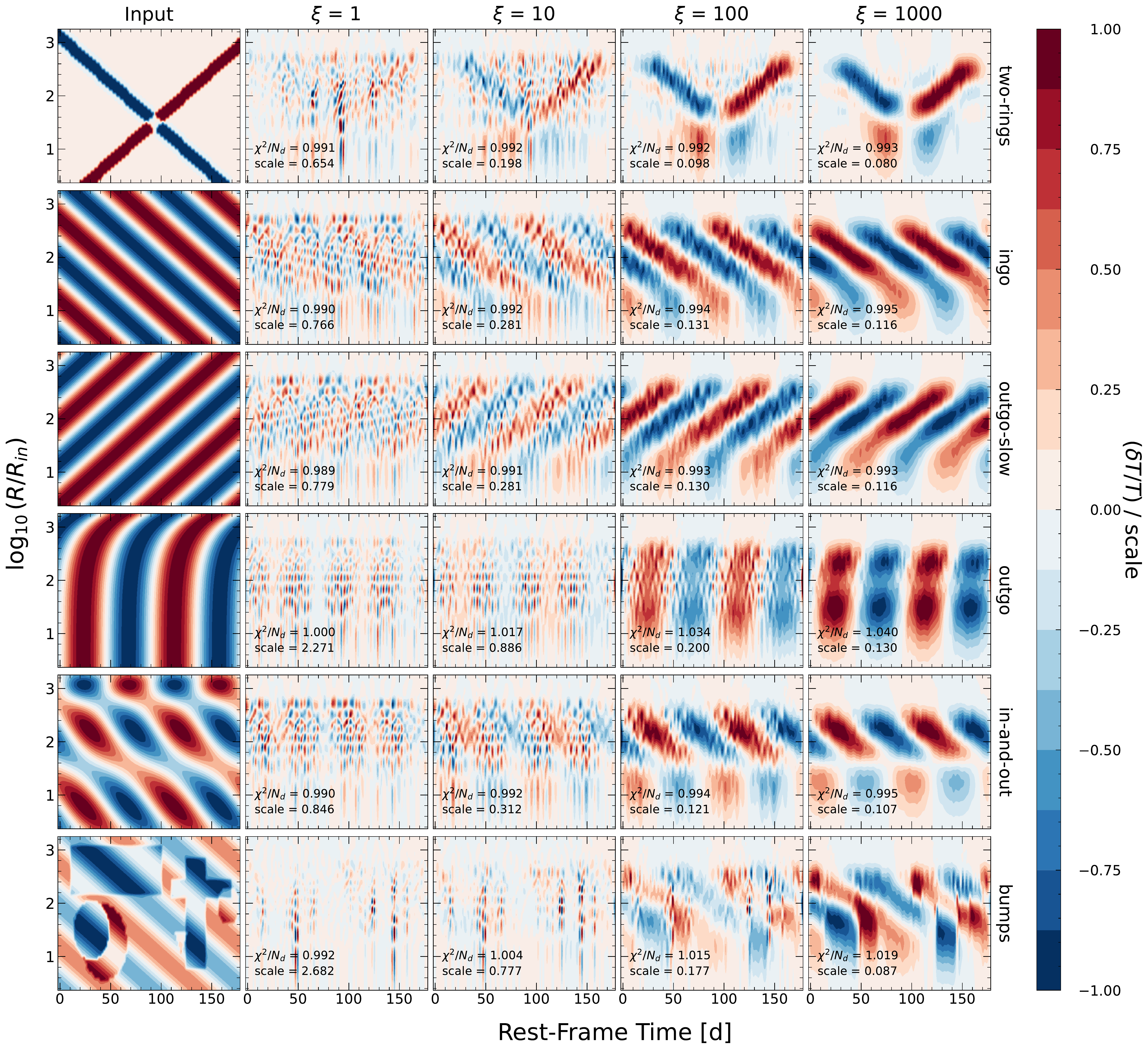}
    \caption{All of the simulated test case scenarios used to test the algorithm with spectra as input, described in further detail in \S \ref{subsec:test_cases}. Each input temperature profile map was constructed with the same parameters, utilizing AGN parameters from NGC5548 and a maximum amplitude of 0.1$T_0(u)$. Each panel displays an output temperature map for a given test case (labeled on the right) and smoothing factor $\xi$ (labeled on top), color coded by $\delta T / T$ relative to the chosen scale. The leftmost column shows the input temperature map for each test case. Like in NK22, we choose to use the 99$^{th}$ percentile of $|\delta T / T|$ as the scale, shown alongside the $\chi^2$ per data point ($\chi^2 / N_d$) for each panel.}
    \label{fig:test_cases}
\end{figure*}

To ensure that this method is robust while using multi-epoch spectra, we test a number of input temperature perturbation maps and compare them against the output (reconstructed) temperature perturbation maps. We start by creating an input temperature map, then use Eqn.~\ref{eqn:flam} to create input spectra, given a grid of observed wavelengths $\lambda$ and observation times $t_d$. We then subtract the mean flux from each epoch to obtain the approximate $\delta F_\lambda (\lambda, t_d)$.  



Observed times and fluxes are then put into the rest-frame of the quasar. We create the $W$ matrix with a given $N_{t_p}$ and $N_u$, used in Eqn.~\ref{eqn:inversion} to reconstruct the temperature perturbation map. We then use $\delta F_{\sigma} = W_{\sigma} \delta T$ to produce the output light curves/spectra, assuming the linear dependence on the temperature perturbations. Thus, for each smoothing factor $\xi$, we obtain $\chi^2 / N_d$ and a scale for the temperature map. 

Each test is performed using the same parameters as input with respect to the resolution of certain variables, error, etc. The spectra are assumed to be observed at a cadence of one day for six months in the observed frame. We use similar spectral range and wavelength sampling as the SDSS spectra for these tests as well, which span $\sim 3000-10000$\AA\, in the quasar rest frame and contain $N_\nu \sim 5000$ data points logarithmically binned in frequency. We use a $u$-grid resolution of $N_u = 50$ in the range defined by the filter kernels, and a $t_p$-grid resolution of $N_{t_p} = 100$. After obtaining the spectra using the input temperature profile, we then add noise relative to the steady-state spectrum $F_{\lambda, SS}$. The uncertainty is chosen from a Gaussian distribution centered at $0.03 F_{\lambda, SS}$ with a standard deviation of $0.005 F_{\lambda, SS}$. The assumed daily cadence and 3\% spectrophotometric precision represent the best possible data quality from ground-based spectroscopic monitoring programs. Furthermore, each test case has a maximum amplitude of $\pm 0.1$ in $\delta T / T_0$. These tests are performed with the same AGN parameters and redshift of NGC5548, to compare with the results from NK22 and span a large range of radii (i.e. $y \in [0.3, 3.1]$).

We utilize many of the same test cases as NK22 (Fig.~\ref{fig:test_cases}): \textit{outgo} is an outgoing wave moving at near the speed of light, resembling a lamppost signal. \textit{ingo} is an ingoing wave whose velocity decreases in magnitude as it approaches the center of the disc. \textit{outgo-slow} is the same as \textit{ingo}, with perturbations traveling outward instead of inward. \textit{in-and-out} is a combination of \textit{outgo} and \textit{ingo} such that the inward traveling perturbations have twice the amplitude as the outward moving ones. \textit{bumps} is an ingoing wave with shapes and spots superimposed, each with an arbitrary amplitude. \textit{two-rings} is the simplest case, with two band-like perturbations, one traveling outward and the other traveling inward at velocities $v \propto u$ like \textit{ingo}. The speed of \textit{outgo} is $\sim 0.5c$, while the speed of \textit{ingo}, \textit{outgo-slow}, and \textit{two-rings} changes as $v \propto u$, such that $v \sim 0.04c$ at $u = 100$.

The results of these test cases vary, though there are some common features of the output temperature maps. Each reconstructed temperature map shows noticeable deviations from the input at small radii, where the emission is not well sampled by the data. Using the more restrictive range on the filter kernels of 0.5 the maximum grants a well-sampled radius range of $y \in [1.3, 2.6]$, effectively excluding the small radii where large deviations from the input map are seen. The smoothing $\xi$ term also has a stronger effect at smaller radii, because the smoothing is conducted in $u$ space while the grid is binned in $\log u$ space. 



The reconstruction for all of the test cases resembles the results in NK22 with photometric light curves. In particular, \textit{two-rings}, \textit{ingo}, \textit{outgo-slow}, and \textit{outgo} are easily recognizable at larger $\xi$. Finer details and perturbations located outside of the well-probed radii are harder to reconstruct, as seen in \textit{outgo, in-and-out}, and \textit{bumps}. The shapes in \textit{bumps} are unresolved and only seen at large $\xi$ as blobs with an an increased amplitude. The shapes in the upper-right and lower left corners of the image are not seen at all. The characteristic turnover at large radii in \textit{outgo} is not seen, though a subtle slant of the vertical structure can be seen. \textit{in-and-out} shows the worst quality reconstruction, resembling an \textit{ingo} pattern due to the range of radii probed. In all cases, a smoothing of $\xi = 100$ provides a balance between recognition of features through smoothing and accurate reconstruction (i.e., $\chi^2 / N_d$ and output amplitude of perturbations).

\begin{figure*}
    \centering
    \includegraphics[width=\textwidth]{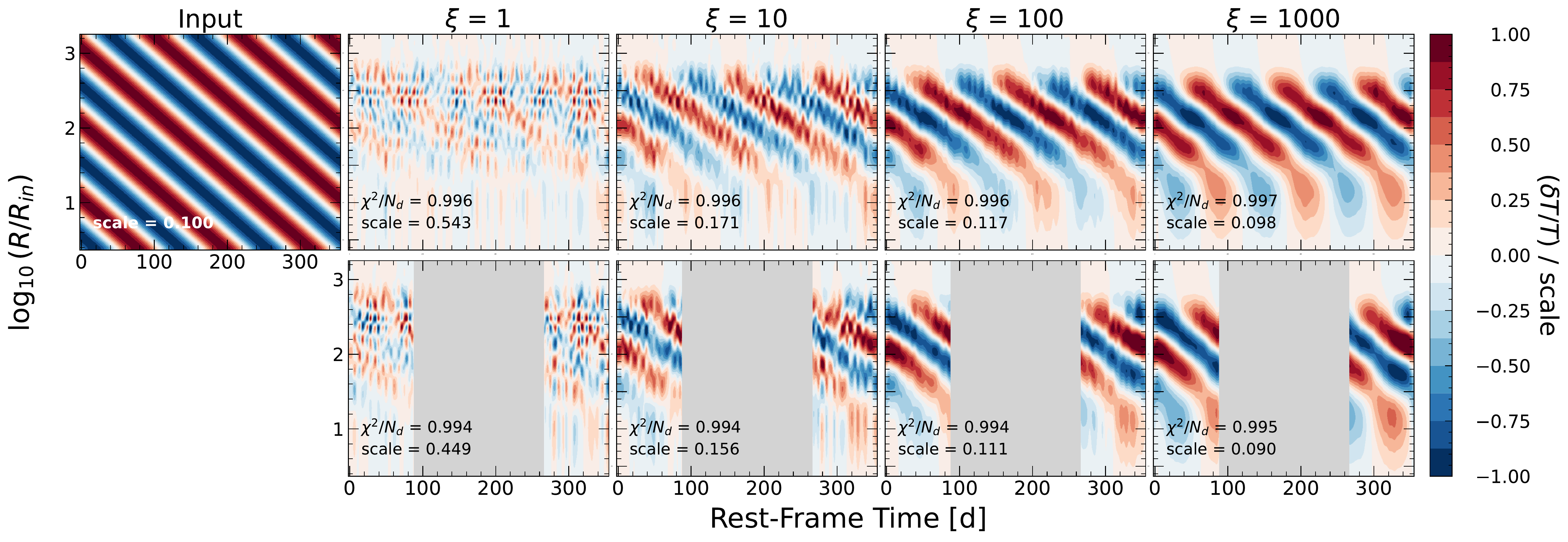}
    \caption{The gap test described in \S \ref{subsec:rm_test}. The input temperature profile map is the \textit{ingo} pattern with a period $\sim 40$ days and an amplitude of 0.1$T_0$. The upper panels represent the output from the algorithm with no gap in observed times. The bottom panel represents the output to the same spectra, but with a gap in observations from 90-270 days, shaded in gray.}
    \label{fig:GapTest}
\end{figure*}

Comparing to NK22, we see that our results resemble the input temperature map at higher $\xi$ within the radius range probed. The boundaries between different parts of the waves are more defined than theirs at larger $\xi$, due to the increase in resolution we have in $\lambda$-space. NK22 describe that many of their temperature maps resemble lamppost-like signals as they increase $\xi$, regardless of the input, while ours resemble the input temperature map even at large $\xi$. At low $\xi$, the method overfits our data due to the increased spectral resolution we have, making the output patterns less obvious. However, this also leads to a $\chi^2 / N_d$ closer to 1 at all $\xi$ as well. One caveat is that NK22's temperature maps span a larger range of radii due to the bands covered by the photometric light curve data they use (spanning from $\sim 1100-9000$ \AA\; in the quasar rest-frame, while our spectra span $\sim 3000-10000$ \AA\; in the quasar rest-frame). However, NK22 also display a smearing at low radii similar to our results, even though they probe smaller radii within the disc, making these features unreliable.

In conclusion, even with the limited spectral range probed by SDSS-like data, it is possible to recover the prominent features in the temperature perturbation map and distinguish fundamental different perturbation patterns (e.g., \textit{ingo} and \textit{outgo}). In addition, the higher spectral resolution is beneficial to resolve radial waves that are otherwise difficult to resolve with coarsely sampled data in wavelength, e.g., photometric light curves. 

\subsection{SDSS-RM Test Cases}\label{subsec:rm_test}


To better understand the performance of temperature perturbation reconstruction using realistic spectroscopic data, we perform tests with sampling properties similar to the SDSS-RM dataset. More general investigations on the reconstruction quality as functions of the spectral resolution and signal-to-noise ratio (S/N), as well as resolutions of the model grid, are presented in Appendix \ref{sec:res_tests}.

Ground-based spectroscopic monitoring usually contains seasonal gaps in the observed epochs. We therefore first test the reliability of this method with respect to gaps in the data, and how well it can reconstruct temperature maps qualitatively. We use an \textit{ingo} pattern as the input with a period $\sim 40$d and a speed $v \sim 0.04c$ at $u=100$. AGN parameters including redshift are the same as for NGC5548 ($z=0.017$). The simulated observations have a cadence of 1 day in the observed frame for a year, and we remove the central six months to create the ``gapped" data. The results in Fig.~\ref{fig:GapTest} show that the reconstruction is resilient to seasonal gaps, producing similar results for both the gapped data and non-gapped data. The perturbation scale is slightly smaller for the gapped data, as the middle section of the temperature map is close to 0, driving down the 99$^{th}$ percentile of the data. The $\chi^2 / N_d$ is lower for the gapped data, as with fewer observations the model overfits the data. The overall structure of the temperature map on either sides of the gap closely resembles that from the input map and the reconstruction without the seasonal gap.

\begin{figure*}
    \centering
    \includegraphics[width=\textwidth]{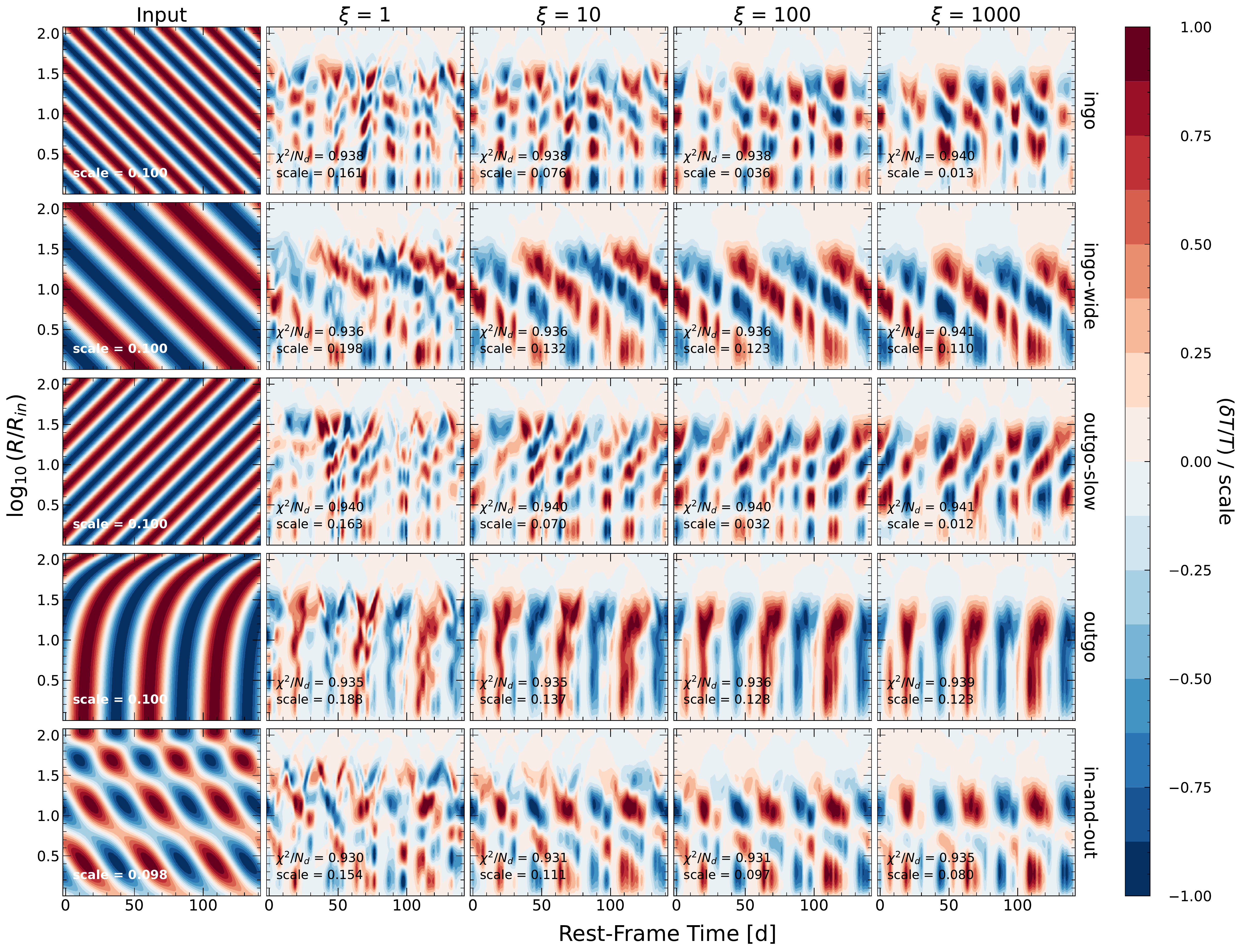}
    \caption{Similar to Fig.~\ref{fig:test_cases}, but with the spectral range, cadence, and AGN parameters of one of the objects in our sample: RMID085. This features only \textit{ingo}, \textit{outgo-slow}, \textit{outgo}, and \textit{in-and-out} with shorter temporal periods than in the previous suite of tests.}
    \label{fig:RMCadenceTest}
\end{figure*}

To test the effects of realistic spectroscopic data on the quality of the fits, we perform the same tests seen in Fig.~\ref{fig:test_cases} with similar properties as the SDSS-RM data set (i.e., spectral coverage, cadence, and S/N). We use the accretion parameters of a particular SDSS-RM target, RMID085, to perform this suite of tests on, as this target is representative of the majority of the SDSS-RM sample in terms of the radii probed by the spectroscopy. The results (Fig.~\ref{fig:RMCadenceTest}) show important deviations from the idealized case shown previously in Fig.~\ref{fig:test_cases}. This is most clearly seen from the output of the \textit{ingo} and \textit{outgo-slow} test cases, where the stripe pattern of the wave becomes fragmented. The output of these test cases display vertical stripes of alternating positive and negative amplitudes, often not showing a definitive straight line like the input, to unambiguously indicate that it is an outward/inward propagating wave with varying speed. In addition, there is a degree of degeneracy between the \textit{ingo} and \textit{outgo-slow} results with low smoothing parameter $\xi$. Both outputs resemble each other very closely with the vertical striped column pattern. However, the different tilting of the stripes becomes more noticeable when increasing the smoothing $\xi$. 

On the other hand, the presence of fast outgoing waves (\textit{outgo}) is much clearer in the output for all smoothing levels $\xi$. NK22 show in their photometric data that for many objects, high-$\xi$ fits produce \textit{outgo}-like patterns. For our spectrocopic light curves, even with low $\xi$ smoothing, this fast \textit{outgo} pattern is noticeable. Though, the combination of \textit{ingo} and \textit{outgo} produces output that resembles \textit{ingo}, at both small and large $\xi$. This presents a degree of degeneracy between \textit{ingo} and \textit{in-and-out} as well. This degeneracy is discussed further in Appendix \ref{sec:in_and_out_test} with Fig.~\ref{fig:InAndOutTest}.

The tests in Fig.~\ref{fig:RMCadenceTest} suggest that with low or moderate-cadence spectroscopic light curves, the reconstruction of the temperature perturbation map will inevitably be degraded compared with the ideal cases shown in \S\ref{subsec:test_cases}, and it is often difficult to unambiguously determine the direction of propagation of slow waves. Nevertheless, even with the SDSS-RM spectroscopic cadence, it is straightforward to distinguish between the slow wave cases (\textit{ingo} and \textit{outgo-slow}) and the lamppost signal case (\textit{outgo}). In addition, the SDSS spectral resolution is sufficient to recover the correct radial frequency of the wave perturbations, as illustrated by the first three input examples in Fig.~\ref{fig:RMCadenceTest} with different radial frequencies.

Another important caveat of the reconstruction using low-to-moderate cadence spectroscopy is the bifurcation of stripes near the top of the map (at large radii). Such ``fanning'' features are most noticeable for results with low smoothing $\xi$. This is an artifact of the reconstruction caused by the smearing effect due to light travel times and sparse-to-moderate cadence of the data. The NK22 method accounts for light-travel-time smearing by averaging over adjacent epochs, which will lead to such structures at large radii if the epochs are sparsely sampled. We illustrate this caveat in detail in Appendix~\ref{sec:smearing}. 



\section{Applications to SDSS-RM quasars} \label{sec:real_data}

\begin{figure}
    \centering
    \includegraphics[width=.5\textwidth]{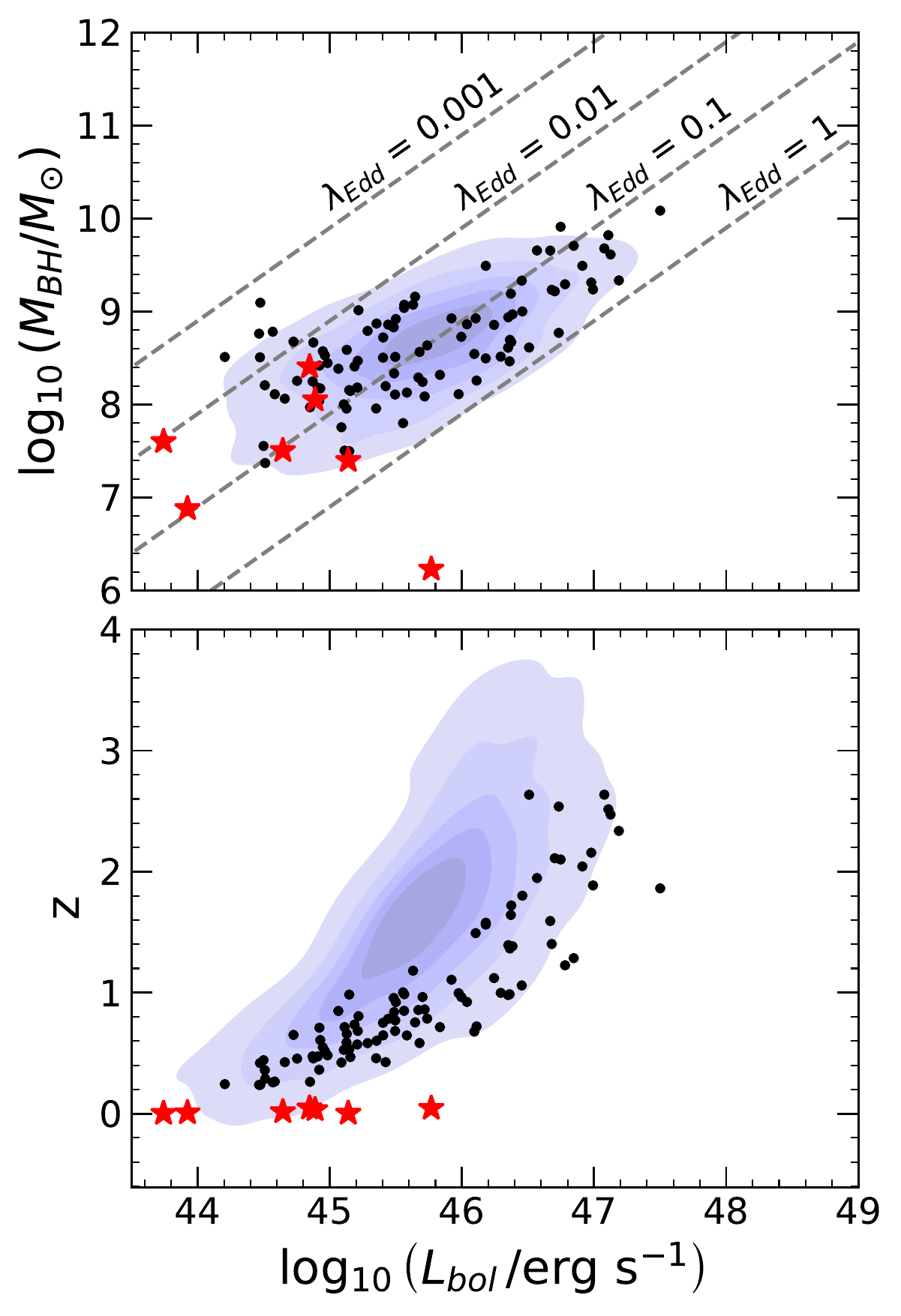}
    \caption{Properties of the $\sim$ 100 quasar sample used from \citet{shen_sloan_2019}, comparing bolometric luminosity $L_{bol}$ to $M_{BH}$ and $z$. Our sample is shown as black points, while contours from the entire SDSS-RM quasar catalog are shown shaded in blue. These contours are constructed to contain [5, 20, 40, 60, 80]$\%$ of the data. The top panel also contains labeled dashed lines of constant Eddington ratio $\lambda_{Edd}$. The sample of local AGNs from NK22 are shown in both panels as red stars.}
    \label{fig:LM_plot}
\end{figure}

\subsection{Data} \label{subsec:getting_data}

To apply the NK22 method to distant quasars, we utilize the multi-epoch spectroscopy from the SDSS-RM project \citep{Shen_etal_2015a}. SDSS-RM is a dedicated multi-object reverberation mapping program that simultaneously monitored $849$ quasars over a broad redshift range of $0.1<z<4.5$ with SDSS spectroscopy during 2014-2020. The primary science goal of SDSS-RM is to measure the time lags between continuum and broad-line emission from distant quasars to infer the size of the broad-line region and derive RM-based black hole masses. But the same data set can also be used to measure continuum lags and accretion disc sizes for high-redshift quasars \citep[e.g.,][]{Homayouni_etal_2019}. There are a total of 90 epochs of spectroscopy over 7 years; 32 epochs were obtained in the first season with an average cadence of $\sim 4$\,days. These SDSS optical spectra cover $\sim 3650-10400$\,\AA\ with a spectral resolution of $\lambda/\Delta\lambda\sim 2000$. 

\citet{shen_sloan_2019} presented properties for the 849 SDSS-RM quasars, including black hole masses and Eddington ratios estimated using the single-epoch virial BH mass estimators \citep{Shen_2013}, as well as measured continuum variability amplitude in the first-season spectroscopic monitoring. We use the compiled continuum variability metrics (SNR2\_C1700, SNR2\_C3000, and SNR2\_C5100) in \citet{shen_sloan_2019} to select 100 most variable quasars. These continuum variability metrics measure the S/N of the detection of intrinsic variability, rather than the variability amplitudes themselves. Given the typical flux measurement uncertainties of SDSS-RM data, intrinsically more variable quasars would on average have higher SNR2 values. 

Fig.~\ref{fig:LM_plot} shows the distributions of these 100 quasars in the luminosity versus BH mass or redshift plane. Compared with the full SDSS-RM quasar sample, these most variable quasars shift to systematically lower redshifts, which is expected given the reduced time dilation to enhance the apparent variability. Nevertheless, these 100 objects still probe a broad range in luminosity and BH masses, and are typical of the high-luminosity quasar population. 

For each spectroscopic epoch, we remove contribution from emission lines and line complexes using the spectral fitting approach outlined in \citet{shen_sloan_2019}. The continuum emission is modeled as a power-law plus a lower-order polynomial. A substantial fraction of these 100 quasars are at $z<1$, where the continuum emission may be contaminated by host galaxy starlight. We use the host-quasar spectral decomposition results in \citet{Shen_etal_2015b} to further subtract host contamination to the accretion disc continuum in these low-redshift quasars. The final accretion disc continuum spectra are used as input to reconstruct the temperature perturbation map for each of the 100 quasars. To best constrain the temperature perturbation map, we focus on the first-season SDSS-RM spectroscopy, which has much higher cadence than subsequent years. The NK22 methodology, however, can be easily applied to multi-season spectroscopy, as illustrated by our test cases in \S\ref{sec:tests}.

\subsection{Results} \label{subsec:results}


We run each quasar in the SDSS-RM sample through the NK22 method, producing an output temperature profile map and output spectra fitted to the input spectra. The data for each object contains the multi-epoch continuum spectra and their errors, and the observed dates of the spectra. The original continuum model is a smooth fit to the data that does not reflect the flux uncertainties per spectral pixel. Therefore, we displace each data point in the smooth continuum model using the uncertainties of the continuum model. This continuum uncertainty is the combination of measurement uncertainties due to flux errors and a systematic fractional uncertainty floor of 5\% from flux calibration \citep{Shen_etal_2015a}, with the latter usually dominating the continuum uncertainties. Each inversion process uses the same parameters of $N_u = 50$ and $N_{t_p} = 100$. While each object has spectra spanning multiple seasons, and the gaps would not have affected the quality of the fits, using a large timespan but a $t_p$ resolution of $N_{t_p}$ would make most of the features within the output map highly unresolved. Additionally, the first season of observing had a much higher cadence than the rest of the seasons, spanning $\sim 100-150$ days, which would be condensed into a small corner of the map if the map spanned all observed times. Therefore, we restrict the map to cover all observations in the first season only, before the first gap in observations. Therefore, we are exploring temperature perturbations on timescales constrained by the first-season of SDSS-RM data (and the redshift of the quasar). 

Admittedly, some quasars have fractional spectral variability well exceeding the linear regime, which represent a population of extreme variability quasars \citep[e.g.,][]{Dexter_etal_2019}. We retain such objects in our sample to reconstruct their temperature perturbation maps. But we caution on the physical interpretation of their results, as the linear perturbation conditions in the NK22 approach are not fulfilled for these extreme variability quasars.

\begin{figure*}
    \centering
    \frame{ \includegraphics[width=.8\textwidth]{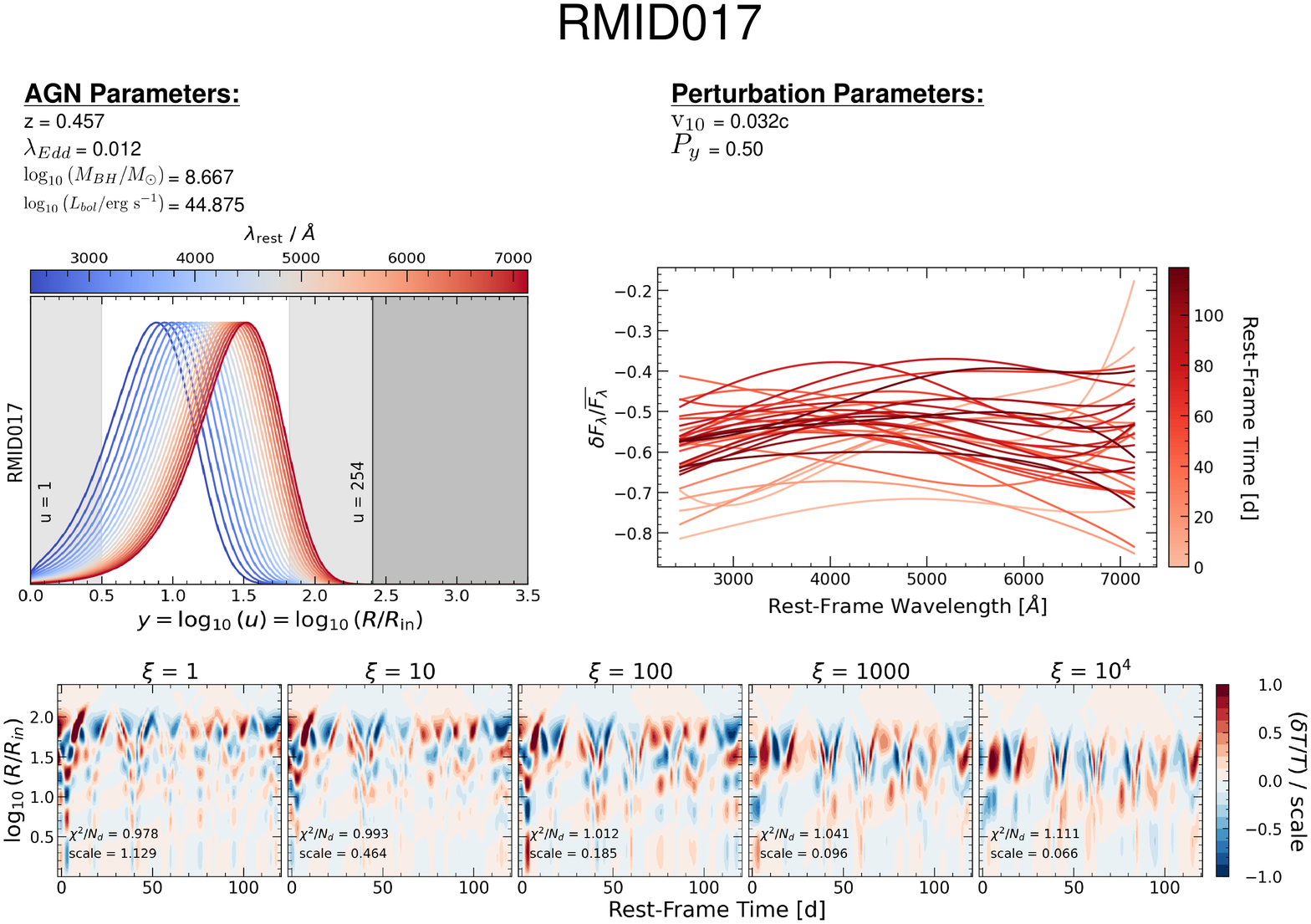} }
    \caption{An example summary file for one of the objects included in the sample of SDSS-RM spectra. This summary file includes: (1) the filter kernels used to get the range of radii probed by the spectra (see Fig.~\ref{fig:kernels}), (2) parameters of the AGN, (3) a figure showing the change in the spectra over time, with the change in spectra relative to the mean (steady-state) spectrum and color-coded by the time of the observation in the quasar rest-frame, (4) the output temperature profile map.}
    \label{fig:summary_file}
\end{figure*}

We compile a FITS table of the continuum spectra and properties for each quasar in the SDSS-RM sample (\url{https://zenodo.org/record/8040692}). In addition, we provide a summary plot for the temperature map reconstruction for the sample, with one example shown in Fig.~\ref{fig:summary_file}. This summary plot contains the parameters of the AGN, the filter kernel used, a plot of the input spectra color-coded by the date of observation, the reconstructed temperature perturbation map, and any parameters to describe perturbations within the map. 


A number of objects in our sample contain outlier spectra for certain epochs, deviating far from the mean (steady state) spectrum. These outlier epochs are most likely caused by systematics in the data processing \citep{Shen_etal_2015a} rather than intrinsic variability. We performed our analysis for these objects both with and without these outlier spectra, to judge their effects on the reconstruction. We find that the reconstructed temperature perturbation pattern is insensitive to this detail. However, the amplitude of the perturbations decreases after the outlier epochs are removed, as the model does not need to account for large jumps in the data between epochs. Given the nature of the smoothing factor $\xi$, the reduction in scale is not as prominent for large $\xi$. We use the results with the outlier rejection as our fiducial temperature perturbation maps.

\begin{figure*}
    \centering
    \includegraphics[width=\textwidth]{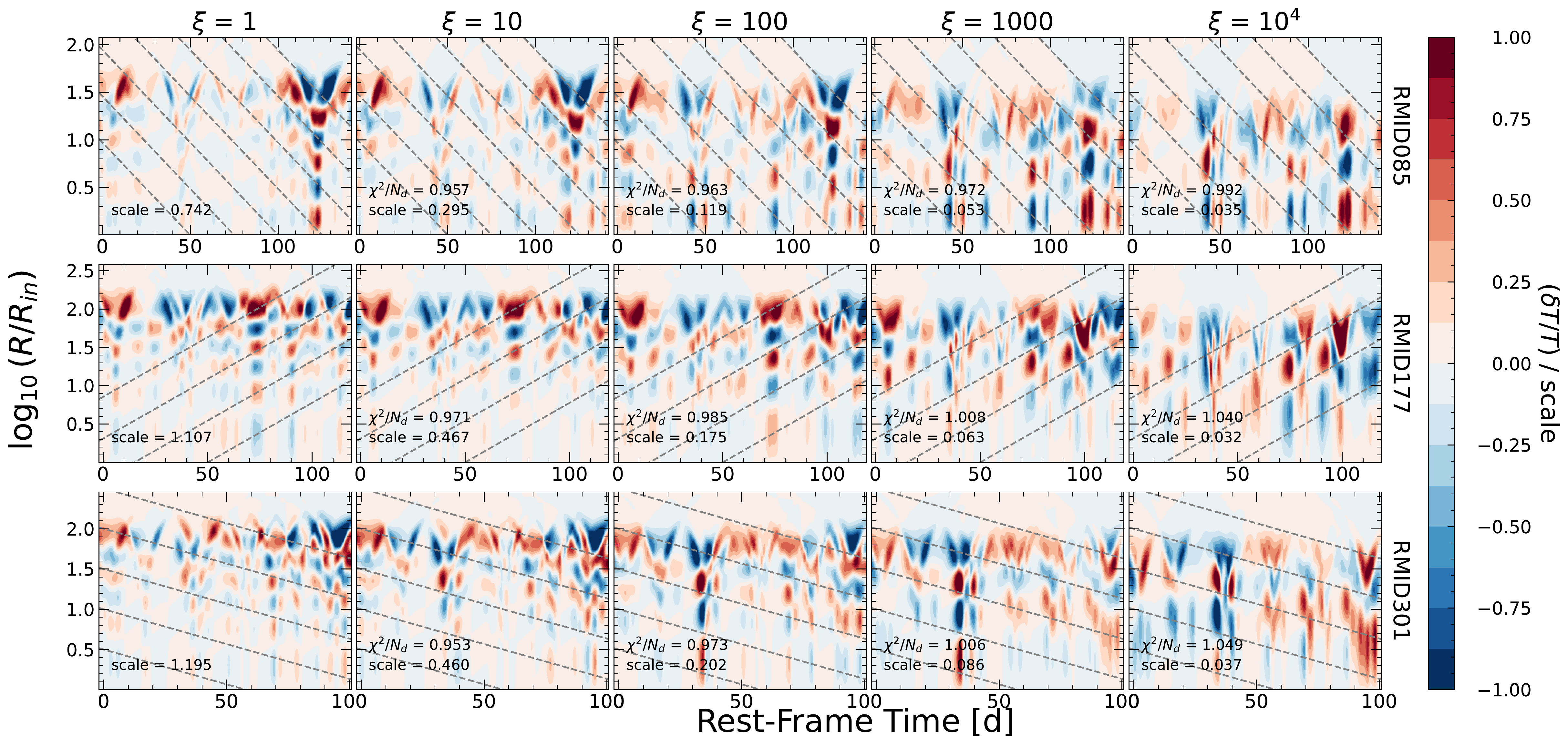}
    \caption{Reconstructed temperature maps for three SDSS-RM quasars. Each temperature map is labeled on the right of the rightmost panel with its name from the SDSS-RM catalog. The maps for each object were visually inspected to determine the direction, speed, and period of their perturbations. Each panel contains dashed lines representing the direction and approximate speed of these perturbations. The width between these lines corresponds to the approximate radial/temporal period of these perturbations as well.}
    \label{fig:RM_output}
\end{figure*}

Each reconstructed temperature map displays one of three visual perturbation patterns: slow outgoing waves (similar to \textit{outgo-slow}), slow ingoing waves (similar to \textit{ingo}), or incoherent perturbation patterns. Most output maps display the columns of alternating positive and negative amplitude, with patches of perturbations, like those seen in Fig.~\ref{fig:RMCadenceTest}. One clear example of this is RMID085, shown in Fig.~\ref{fig:RM_output}, which displays resemblance to the \textit{ingo} pattern, though it is visually similar to an \textit{outgo-slow} pattern as well (see discussions on this ambiguity in \S\ref{subsec:rm_test}). For all objects, the presence of perturbations is subtle at low $\xi$, only becoming clear once reaching large $\xi \gtrsim 10^3$. For the majority of our sample we find no evidence of any coherent pattern of fast outgoing waves in the output temperature maps. This is somewhat different than NK22 for local AGNs, where large $\xi$ often leads to lamppost-like (\textit{outgo}) behavior due to heavy smearing of the temperature perturbations in the radial direction. We have tested with downsampling the spectral resolution to mimic the photometric light curves in NK22, but still find no lamppost temperature perturbation patterns for most SDSS-RM quasars with $\xi$ as large as $1000$. We suspect that this difference is likely caused by the differences in the target samples studied in both studies, i.e., different BH masses and luminosities (see Fig.~\ref{fig:LM_plot}).

Most quasars show stochastic variability in their temperature maps, without strong coherent visual patterns. However, a slow ingoing/outgoing pattern can be recognized for many of them over limited periods (several examples are shown in Fig.~\ref{fig:RM_output}). It is possible that superpositions of incoherent waves can produce such visual patterns. Alternatively, unknown systematics in the continuum spectra could also contribute to the stochastic pattern seen in the reconstructed temperature perturbation map. When comparing our maps with those in NK22 for local AGNs, we notice that the reconstructed temperature maps in NK22 also have many stochastic features. Thus these incoherent patterns seen in our sample are likely intrinsic rather than due to systematics.



Even at smoothing factors as large as $\xi = 10^4$, we do not see any structures resembling lamppost-like waves. In our test cases shown in \S\ref{sec:tests}, we see that lamppost signals would appear at large $\xi$ if they were present in the spectra, regardless of the cadence, spectral range, or resolution in $t_p$. It may be possible that there are low-amplitude lamppost-like perturbations superimposed on top of the dominant slow ingoing/outgoing perturbations. However, we test this using simulated combinations of \textit{ingo} and \textit{outgo}, and find that if there were any low-amplitude lamppost-like waves, they would be visible at large $\xi$ (see Appendix \ref{sec:in_and_out_test}). The fact that these lamppost patterns are not seen at large $\xi$ for SDSS-RM quasars indicates that these ingoing and outgoing perturbations are real, and not systematics from the reconstruction.

We do observe systematic patterns that have been amplified with respect to the initial tests in \S\ref{subsec:test_cases}. The most prominent systematic feature in the temperature maps is a ``fanning out" seen at large radii, near the edge of the range of probed radii. This is seen in the test cases with the observational parameters of RMID085 (Fig.~\ref{fig:RMCadenceTest}) at large radii for low $\xi$. This can be seen in Fig.~\ref{fig:RM_output} as well, with the fanning out occurring at the upper limit of the radius range probed for all $\xi$, for each object. This fanning is seen on top of vertical, striped columns for almost all objects. A single ``fan" can encompass the entire baseline of observations for some objects. In Appendix \ref{sec:smearing}, we argue that this fanning pattern is due to the smearing effect introduced in \S\ref{sec:methods}. We do not observe fanning for the idealized test cases in \S\ref{subsec:test_cases}, because the spectra are too highly sampled in $t_d$ space. Each observation at a given $t_d$ influences a range of parameter times $t_p$ at a given radius $u$. This range of times increases as $u$ increases, causing this fan pattern. If the $t_d$ resolution is low, these fans are noticeable from the widely separated individual observations at large radii. At higher resolutions, these fanning patterns overlap at large radii, adding up and making their effect on the temperature map minimal (discussed further in Appendix \ref{sec:smearing}).


Using $\xi=100$ as the most accurate smoothing factor for the scale of the temperature map (see Appendix \ref{sec:res_tests}), most objects have perturbations $\sim 0.1 T_0$. This is both consistent with the results from NK22, and with the linear regime presented in their method. Though, there is uncertainty in the suggested temperature perturbation amplitudes in AGN: \cite{dexter_quasar_2010} find that describing temperature fluctuations as a random walk produces an amplitude $\sim 40\%$ when comparing to flux observations of AGN. \cite{hirose_turbulent_2009} produce temperature variations with an amplitude $\sim 2T_0$, while general relativistic magnetohydrodynamic (GRMHD) simulations by \cite{fragile_global_2007} and \cite{mckinney_stability_2009} result in amplitudes of $10-20\%$, though utilize thick discs and do not consider radiation dynamics.

Many SDSS-RM quasars have maps that contain visual patterns similar to \textit{ingo} and \textit{outgo-slow} that can be traced by eye in terms of direction and speed, as well as radial/temporal period. These more obvious perturbations have speeds $ v_{10} \sim 0.02c$ at $u=10$, and radial periods $P_y \sim 0.5$ dex. The logarithmically radial periods we find are shorter than in NK22, and the speeds we find are faster by a factor $\sim 10$. This discrepancy can be due to both the degeneracy in different types of outgoing/ingoing perturbations due to the fragmentation from temperature map reconstruction, and the properties of the AGN we study. The range of radii that the NK22 sample probes is on average larger than the range for our sample. Thus, we may only be seeing short-term perturbation motion for an overall larger perturbation pattern within the disc. Extending to a larger region in the disc may show that the perturbations we observe are stochastic noise relative to larger, more significant perturbation patterns. 

In terms of physical timescales, these perturbations have periods $\sim 40$ days, similar to the typical orbital timescale at a radius $u \sim 10$ for $M_{BH} = 2 \times 10^8 M_{\odot}$. If the orbital timescale were much quicker than the period of the wave-like perturbations, the perturbation may be smoothed out from the rotation of the disc. The fact that these two timescales match at similar radii suggest that instabilities could develop on the orbital timescale that give rise to outgoing/ingoing wave-like perturbations. In addition, these perturbations travel coherently over intervals of $\sim$ months, also similar to the dynamical time at these inner radii for a typical $M_{BH}$ in our sample.

These perturbations are also of constant width in $y$-space, growing larger as they travel farther out into the disc. This behavior likely follows from viscosity, thermal energy transfer, and rotation causing the perturbation to lose energy and diffuse into the disc. However, these perturbations generated within the disc are likely more complicated than the simple linear perturbations from \textit{ingo} and \textit{outgo-slow}. NK22 find perturbations that move inward and outward through the disc at different times within their temperature maps. These traveling perturbations may also change in amplitude as they travel due to loss of energy. Our reconstructed temperature perturbation maps are roughly consistent with these expectations, although more quantitative constraints are difficult to derive given the nature of this approach and assumptions therein. 


An important caveat to these results is possible contamination to the continuum emission from the accretion disc. One of such sources of contamination is internal reddening on the SDSS-RM sample. For our chosen sample, we do not have the ability to distinguish host contamination and internal dust reddening from the continuum. The method of continuum extraction described in \citet{Shen_etal_2015a} have shown to produce consistent results for host-galaxy subtraction, without any bias from reddening \citep{Yue_etal_2018}. Omitting reddening from analysis has shown to overestimate AGN luminosities by a factor of nearly 4 in the optical and 10 in the ultra-violet \citep{gaskell_case_2017}. We have performed tests by adding reddening to our spectra, such as with the reddening curve observed by \citet{gaskell_agn_2007}, and found that the output temperature maps are similar, but temperature variations are suppressed at small radii. At extreme reddening, the output temperature map will concentrate all variation at the largest radii probed, which is not seen in our results. In addition, fits with more reddening produced much higher $\chi^2 / N_d$ values, indicating that if reddening were prominent in our sample, we would see so in the quality of our fits. Accounting for reddening in the SDSS spectra also does not significantly alter the output temperature maps, and produces poorer fits. Thus, while the effects of reddening do not affect our sample significantly, our temperature maps place an upper bound on the temperature variations within the accretion disc. Additionally, it has been shown that diffuse BLR emission is the dominant source of the recent accretion disc time delay measurements \citep{netzer_continuum_2022}, which can contaminate continuum emission. The NK22 method assumes that all variable emission comes from a thin accretion disc. Variable emission from the BLR, as well as internal reddening/host-galaxy emission, can be included in this model as a possible extension of this work.

\section{Conclusions} \label{sec:conclusion}

The nature of the origin of AGN stochastic variability is uncertain, both in the mechanisms producing perturbations within the disc, and mechanisms that affect the propagation of these perturbations and create the light curves we observe. The lamppost model has been invoked to describe disc variability as fast, outgoing wavelike temperature perturbations traveling through the disc at near the speed of light, originating near the SMBH. There has also been evidence of slower perturbations traveling both inward and outward throughout the disc, originating at various disc radii. NK22 have recently developed a method to utilize AGN light curve data to produce temperature perturbation maps of the accretion disc as a function of time. 

In this work, we perform an extension of the novel method described in NK22 to utilize input multi-epoch spectroscopic light curves, and apply to the spectroscopic monitoring data from the SDSS-RM project \citep{Shen_etal_2015a}. We perform a series of tests using certain idealized input temperature perturbation maps: (i) using accretion parameters from a nearby AGN (NGC5548) with idealized observational cadences, spectral ranges, and uncertainty, and (ii) using accretion parameters typical of distant SDSS-RM quasars, with the observational cadence, spectral range, and uncertainty representative of the SDSS-RM sample. The results from these tests are as follows:

\begin{enumerate}
    \item[$\bullet$] Our high spectral resolution produces reasonably good temperature perturbation maps in the range of radii probed by the data, even though the temporal resolution (i.e. cadence) is low. Comparing to NK22, we find similar $\chi^2 / N_d$ and temperature perturbation amplitudes in our suite of tests.
    
    \item[$\bullet$] There is a degeneracy between both the slow, ingoing wave (\textit{ingo}) and the slow, outgoing wave (\textit{outgo-slow}) in recovered temperature maps for irregular or low cadences of the monitoring spectroscopy. Both input patterns produce vertical columns with stripes of alternating positive and negative perturbation amplitudes. These columns produce fragmented waves that look similar for both ingoing and outgoing perturbations.
    
    
    \item[$\bullet$] The smearing term accounting for light travel times across the accretion disc produces a systematic artifact on the output temperature maps at large radii if the cadence of observations is moderate to low. At large radii near the edge of the range probed by the observed wavelengths, there is a pattern of bifurcation seen at the top of the aforementioned striped columns. These ``fanning out" patterns are due to the lack of averaging out of the smearing term at large radii for low-cadence observations. At higher cadences, these ``fans" average out over many epochs and largely disappear.
    
\end{enumerate}

These tests demonstrate that with high-quality spectroscopic monitoring data, especially those with high cadences (e.g., every 1--2 days), high-fidelity temperature perturbation maps can be reconstructed from such data. Seasonal gaps will not introduce systematic biases in the reconstruction. With reduced cadences (e.g., similar to those for SDSS-RM quasars), the quality of the temperature map reconstruction is degraded. However, it is still straightforward to differentiate between the slow-moving waves and the fast-moving lamppost patterns.
 
We then perform this analysis using spectroscopic monitoring data for 100 of the most variable quasars from the SDSS-RM \citep{Shen_etal_2015a} sample. These quasars have been monitored for multiple seasons, and we use the first-season data with a high cadence of $\sim 4$ days to study variability over days to months timescales. For each quasar, we generate the reconstructed temperature perturbation map given the properties of the quasar and the spectroscopic monitoring data. The main results for the SDSS-RM sample are the following:

\begin{enumerate}
    \item[$\bullet$] The vast majority of SDSS-RM quasars display incoherent stochastic variability in their global temperature perturbation maps. But there are often patches of regions that display clear evidence of slow, inward- or outward-moving perturbations, similar to \textit{ingo} and \textit{outgo-slow}. There is no clear evidence of fast, outgoing lamppost-like signals in any of the reconstructed temperature maps. While the cadence of the SDSS-RM data is insufficient to well resolve the light travel time across large disc radii, the characteristic vertical stripes from the lamppost model at small disc radii are not seen.
    
    \item[$\bullet$] For visually recognizable slow wave patterns, the typical perturbation amplitude is $\delta T/T_0\sim 10\%$, and the typical temporal frequency of the wave is $\sim 40$ days (in quasar rest-frame). This average timescale is similar to the orbital timescale of the disc at $R \sim 10R_{\rm in}$ for typical BH masses ($\sim 10^8\,M_\odot$) of the SDSS-RM sample. These perturbations have a speed $\sim 0.02c$ at $R=10R_{\rm in}$, faster than the speeds seen in NK22 by a factor $\sim 10$. Of course, these wave speeds are approximate at best, since it is difficult to measure the exact speed in the reconstructed temperature perturbation maps. 
\end{enumerate}

Our results are consistent with the findings in NK22. Temperature perturbations in AGN and luminous quasar accretion discs are not dominated by the lamppost signal, indicating disc instabilities are the main driver for accretion disc temperature perturbations. However, as pointed out by NK22, when measuring the flux variability, contributions from slow-moving temperature perturbations are substantially more suppressed due to radial averaging, resulting in a more prominent lamppost signal in the light curves. Over longer observing baselines, the flux contribution of the slow-moving temperature perturbations becomes more important and dilutes the lamppost signals. Furthermore, there are many cases where the temperature perturbations are completely dominated by incoherent slow-moving waves, which means it would be difficult to use the RM technique to measure continuum lags and to infer the disc sizes and temperature profiles. Even in cases with coherent slow-outgoing waves, the interpretation of the measured continuum RM lags would be significantly complicated since the perturbations are not propagating at the speed of light, as typically assumed in continuum RM studies.  

Our general conclusion from this study is that the NK22 approach is a very promising tool in further constraining accretion disc variability with upcoming photometric and spectroscopic monitoring of AGN and quasars, e.g., with the Vera C. Rubin Observatory Legacy Survey of Space and Time \citep{Ivezic_etal_2019} and the SDSS-V Black Hole Mapper program \citep{Kollmeier_etal_2017}. In Appendix \ref{sec:res_tests} we demonstrate the utility and caveats of LSST-like light curves on constraining the disc temperature fluctuations (Fig.~\ref{fig:LSST_Test}), which are expected to be an important data set to study accretion disc variability \citep[e.g.,][]{Kovacevic2022}. Such time series data with high cadences, broad spectral coverage and adequate S/N would enable reliable reconstruction of the accretion disc temperature perturbation map for large samples of AGNs and quasars. The increased baselines will also enable the exploration of the emergence and propagation of temperature fluctuations over extended periods of time.

\section*{Acknowledgments}

We thank the anonymous referee for their useful and insightful  comments on this work. We thank Jack Neustadt and Chris Kochanek for help with the implementation of their method, as well as useful comments on the draft. Z.S. acknowledges support from the Center for AstroPhysical Surveys (CAPS) at the National Center for Supercomputing Applications (NCSA), University of Illinois Urbana-Champaign, and Y.S. acknowledges partial support from NSF grant AST-2009947. 

This research made use of Astropy, a community-developed core Python package for Astronomy \citep{2018AJ....156..123A, 2013A&A...558A..33A} This research made use of SciPy \citep{virtanen_scipy_2020} This research made use of NumPy \citep{harris_array_2020} This research made use of matplotlib, a Python library for publication quality graphics \citep{hunter_matplotlib_2007}. 

Funding for SDSS-III has been provided by the Alfred P. Sloan Foundation, the Participating Institutions, the National Science Foundation, and the U.S. Department of Energy Office of Science. The SDSS-III web site is \url{http://www.sdss3.org/}. 

SDSS-III is managed by the Astrophysical Research Consortium for the Participating Institutions of the SDSS-III Collaboration including the University of Arizona, the Brazilian Participation Group, Brookhaven National Laboratory, University of Cambridge, Carnegie Mellon University, University of Florida, the French Participation Group, the German Participation Group, Harvard University, the Instituto de Astrofisica de Canarias, the Michigan State/Notre Dame/JINA Participation Group, Johns Hopkins University, Lawrence Berkeley National Laboratory, Max Planck Institute for Astrophysics, Max Planck Institute for Extraterrestrial Physics, New Mexico State University, New York University, Ohio State University, Pennsylvania State University, University of Portsmouth, Princeton University, the Spanish Participation Group, University of Tokyo, University of Utah, Vanderbilt University, University of Virginia, University of Washington, and Yale University.

\section*{Data Availability}

We provide all data at \url{https://zenodo.org/record/8040692}, and all supplemental figures, animations, and the implemented python code of the NK22 method at \url{https://github.com/Zstone19/TempMap}.





\bibliographystyle{mnras}
\bibliography{sample631,TempMaps,refs} 




\begin{appendix}

\section{The Smearing Function}\label{sec:smearing_deriv}

\begin{figure}
    \centering
    \includegraphics[width=.5\textwidth]{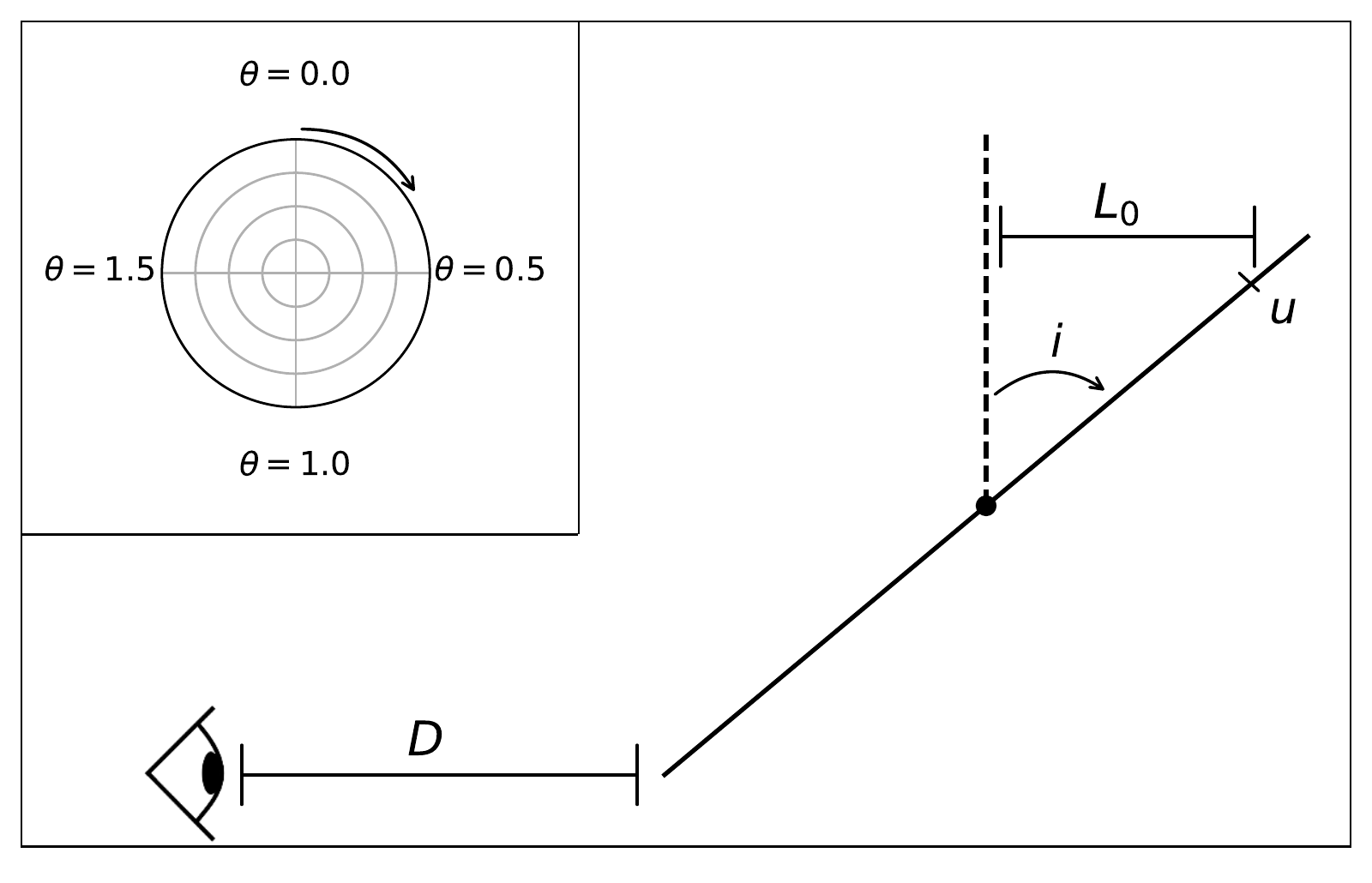}
    \caption{A schematic diagram of the geometry of the accretion disc-observer system for constructing the smearing term, described in Appendix \ref{sec:smearing_deriv}. The inset plot in the top left displays how we define the polar coordinate $\theta$ on the accretion disc.}
    \label{fig:smearing_diagram}
\end{figure}

NK22 present formalism to represent the smearing that occurs in observed AGN light curves  due to the time delay across different regions of the accretion disc. Light observed at a given time (in the observer's frame) is the combination of light emitted from different parts of the disc at different times (denoted as ``$t_p$'' for parameter or proper time in the disc reference frame), which would ``smear out'' variability and emission overall. We parameterize the azimuthal position on the disc using the polar angle $\theta$ in units of $\pi$, which is 0 at the top of the disc, 0.5 to the right of the disc, 1 at the bottom, etc. 
To begin, we lay out the general geometry of the system (Fig.~\ref{fig:smearing_diagram}): an inclined disc with respect to the observer's line of sight ($i$ is the angle between the LOS and the disc normal) located at a distance $D$ from the observer (at the near-side of the disc). There is a characteristic length $L_0$ that a photon emitted at a radius $u$ on the disc needs to travel to the center of the disc in the LOS direction: $L_0 = u R_{in} \sin(i)$. Converting this to a timescale: $t_0 = \frac{u R_{in} \sin(i)}{c}$. For simplicity, all times are assumed to be in the reference frame of the AGN.

After being emitted at a time $t_p$, light will arrive at the observer at time $t_d = D/c + t_0\left[ 1 + \cos(\pi\theta) \right]$. We redefine the emitted time (i.e. the ``parameter time") relative to when emission at the center of the disc is observed, which produces: $t_d = t_p + t_0 \cos(\pi\theta)$. Rearranging, $\theta = \frac{1}{\pi} \cos^{-1}\left( \frac{t_d - t_p}{t_0} \right)$.

To measure the effect of the smearing, we require a function that tells us how to weight $t_p$ values for a given $t_d$ and $u$. In other words, we need the range of $t_p$ that influence a given $t_d$ and $u$. To do this, we differentiate $\theta$:

\begin{equation}
d\theta = \frac{dt_p}{\pi t_0}\left[ 1 - \left( \frac{t_d - t_p}{t_0} \right)^2 \right]
\end{equation}

This gives us the range in $\theta$ (i.e. the area on the disc) that influences a given measurement for a given radius $u$. This is one way to weight the $t_p$ for a given $t_d$, $u$. However, in practice, we only sample $t_p$ on a finite grid, with spacing $\Delta t$. To account for the mismatch in $t_p$ and $t_d$, we interpolate the previous weighting function $d\theta$ to $t_d$. To perform this interpolation, we convolve $d\theta$ with a triangle function centered on $t_p$ with width $\Delta t$:

\begin{equation}
f(u, t_p, t_d) = \int \left[ 1 - \left| \frac{t - t_p}{\Delta t}\right| \right] \left[ \frac{1}{\pi t_0} \left( 1 - \left( \frac{t - t_d}{t_0} \right)^2 \right)^{-1/2}\right] \ dt
\end{equation}

Care must be taken in performing this integral, namely because the two functions used are not defined for all $t$. The two regions of the integral (i.e., the left and right side of the triangle function) are defined by Eqn.~\ref{eqn:int_bounds} as $[t_1, t_2]$ and $[t_3, t_4]$. Performing this integral produces the form seen in Eqn.~\ref{eqn:smearing_tot} using the smearing functions defined in Eqns.~\ref{eqn:G_term1} and \ref{eqn:G_term2}.

There are several caveats to this formulation that could affect the output of the smearing function. Firstly, for the convolution to be finite, the two functions (i.e., $d\theta$ and the triangle function) must overlap. Additionally, the left side of the integral (i.e., from $t_1$ to $t_2$) must be 0 if the left side of the triangle function does not overlap with $d\theta$, and similarly for the right side. All conditions and their effects on the smearing term are shown in Table \ref{tab:smearing}.

\begin{table}\label{tab:smearing}
\centering
\caption{Additional Constraints on the Smearing Function $f(u,t_p,t_d)$}
\small
\begin{tabular}{|c|c|}
\hline
Condition & Effect \\
\hline
$t_p + \Delta t < t_d - t_0$ & $f(u,t_p,t_d) = 0$ \\
$t_p - \Delta t > t_d + t_0$ & $f(u, t_p, t_d) = 0$\\ 
$t_p < t_d - t_0$ & $t_1 = t_2 = 0$ \\
$t_p > t_d + t_0$ & $t_3 = t_4 = 0$ \\
\hline
\end{tabular}
\end{table}

\section{Resolution Tests}\label{sec:res_tests}


\begin{figure*}
    \centering
    \includegraphics[width=\textwidth]{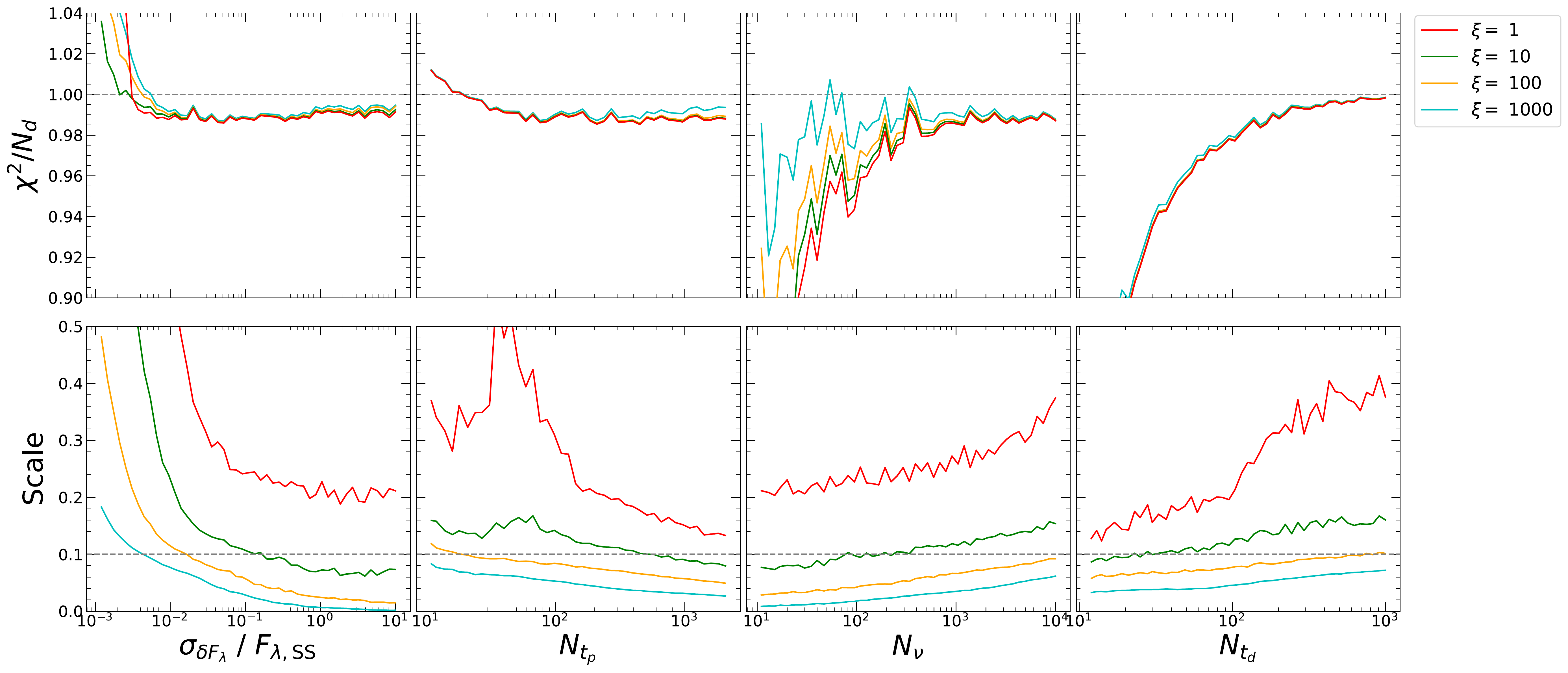}
    \caption{The result of our suite of tests for different resolutions of $t_p, t_d$, and $\lambda$, and relative error with respect to the steady state spectrum $F_{\lambda, SS}$. These tests were performed with the same parameters as the tests shown in Fig.~\ref{fig:test_cases}, described in \S\ref{subsec:test_cases}, using the \textit{ingo} test case temperature map. Each inversion produced a result for $\xi=1,10,100,1000$, plotted in different colors on each plot. The evolution of $\chi^2 / N_d$ (top) and the scale (bottom) of the map are shown for each parameter. $\chi^2 / N_d$ = 1 and the input temperature map scale (0.1) are labeled with dashed gray lines in each panel.}
    \label{fig:chi2}
\end{figure*}

We test the effect of the resolution of $t_d$, $t_p$, $\lambda$, and the relative error of the input spectroscopic light curves on the quality of the fits. The quality of the inversion is measured both by $\chi^2 / N_d$ and the output perturbation amplitude (i.e. the scale). We performed this test using the \textit{ingo} test case, with the same parameters described in \S \ref{subsec:test_cases} and a period of $\sim 40$d (Fig.~\ref{fig:chi2}). Among these four tested parameters, $t_p$ is related to the model setup while the other three are related to the data quality.

The most obvious result from these tests is the dependence of $\chi^2 / N_d$ on the cadence of observations, relative to the characteristic length of the signal received from the perturbation. There is a clear trend of $\chi^2 / N_d$ approaching 1 as the cadence becomes shorter (i.e. the number of observed times $N_{t_d}$ increases) for all smoothing factors. This is to be expected as more observations produces more constraints in the inversion and more data to use when inverting, resulting in a better fit. The dependence of the output amplitude on the cadence is the inverse, growing at a quicker rate for shorter cadences. This can be seen in Fig.~\ref{fig:NtdTest}, showing the output of the temperature maps for different cadences. For large cadences, perturbations introduced at each observed time are smeared out over a range of parameter times $t_p$. As the number of observed times increases, these ranges for each time overlaps and leads to larger perturbations in general. Though, as more data is introduced with the addition of more $t_d$ points, the $\xi=1,10$ fits overfit the data more, leading to a noisier temperature map with less of a pattern. The scale of the input perturbations is best matched by the output maps at $\xi=100$ for almost all of the tests in the default range we use for them.

The scale and $\chi^2 / N_d$ display clear dependencies on the spectral resolution ($N_\nu$). The algorithm overfits the data at low spectral resolution, but produces fits with $\chi^2 / N_d \sim 1$ as the resolution increases, similar to the dependence on cadence. The scale shows similar behavior as well, increasing as the resolution increases. This behavior is analogous to the behavior seen with cadence as well - for low resolution spectra, perturbations introduced at each wavelength apply to a range of radii (governed by the filter kernels in \S\ref{sec:methods}). Therefore, the sparsity of spectral data in low resolution examples show more smearing in the radial direction. The effect is less than in $t_d$-space, as there is also an explicit smearing term involved in the inversion. This smearing comes from the weighting of different radii for each band, defined in Eqn.~\ref{eqn:delta_flam}. This explains why the increase in scale is slower than for the cadence test. Indeed, Fig.~\ref{fig:NnuTest} showing output temperature maps for different spectral resolutions displays overfitting at $\xi=1,10$ for high resolution spectra. This effect wanes as the smoothing increases, as the $\xi$-smoothing dominates this "spectral smoothing". 

The quality of fits as the resolution in $t_p$-space changes matches the behavior seen in NK22: as the resolution increases, both the scale and $\chi^2 / N_d$ decrease. They attribute the decrease in scale to the fact that perturbations can be spread out over more times $t_p$, rather than concentrated at a single (or a few) times. These spread-out fluctuations are then cause the $\chi^2$ to more easily be minimized, as their scale has decreased. However, the change in $\chi^2 / N_d$ is rather small, meaning the effect on minimization of $\chi^2$ is not that significant. The effect of the amount of error in the input spectra resembles that of $t_p$ resolution. As the error increases, $\chi^2 / N_d$ decreases. All equations used to minimize $\chi^2$ are relative to the error, meaning that large errors cause the spectra to be fit easier. In essence, there is more room for error in the $\chi^2$ fitting. Looking at Fig.~\ref{fig:ErrTest} we can see that for lower errors ($\sim 10^{-3} F_{\lambda, SS}$), the data is overfit for all smoothings because of the small error. As the error increases, the smoothing becomes more apparent. The scale decreases as the error increases, as the perturbations from a single time are smoothed out across multiple times, similar to the resolution in $t_p$-space. This is also due to the fact that as the error increases, the nature of the smoothing factor $\xi$ causes the output to tend towards no perturbations with little constraint. With large error, there is little constraint on the shape of the model, so the algorithm produces highly smoothed out temperature maps, with smaller amplitude perturbations.

\begin{figure*}
    \centering
    \includegraphics[width=\textwidth]{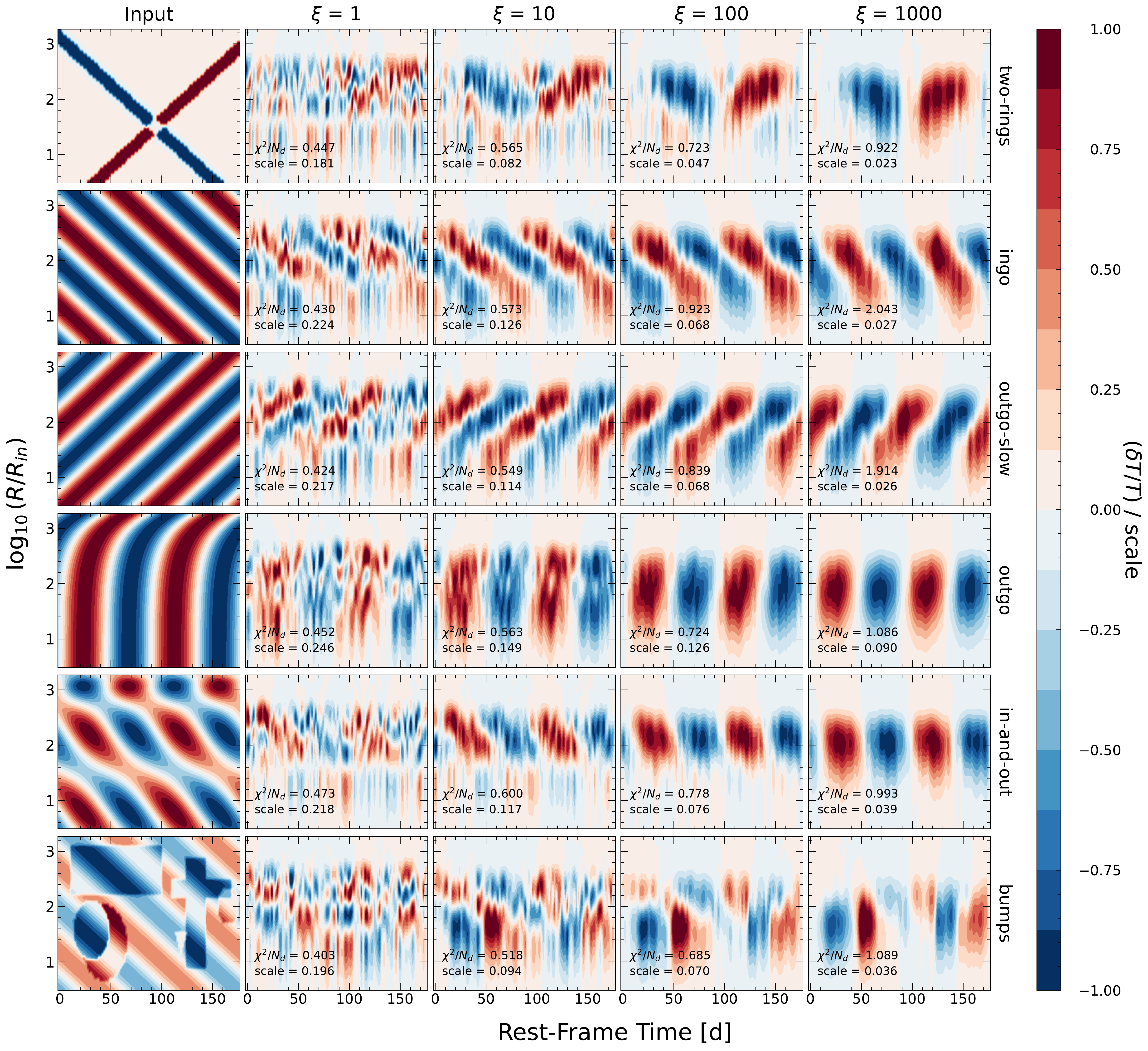}
    \caption{The same suite of tests described in Fig.~\ref{fig:test_cases}, using a cadence of 2 days and spectral sampling in the ugrizY bands, typical of observations from the Vera C. Rubin Observatory Legacy Survey of Space and Time \citep{Ivezic_etal_2019}.  }
    \label{fig:LSST_Test}
\end{figure*}

In addition to the test of generic resolution for our idealized set of observations in Fig.~\ref{fig:test_cases}, we also perform a test with the same NGC 5548 parameters and input temperature maps for high-cadence Vera C. Rubin Observatory Legacy Survey of Space and Time \citep{Ivezic_etal_2019} observations (Fig.~\ref{fig:LSST_Test}). We use a cadence of two days and sample the light curves in the ugrizY bands. \citet{Kovacevic2022} show that this short cadence provided by the Deep-Drilling Fields (DDFs) is optimal for accretion disk lag studies, and will be able to constrain time lags between the LSST bands for thousands of quasars. As there is much less data to fit, the data is overfit for low $\xi$ in all cases. $\xi=1000$ represents the best fit to the data in terms of $\chi^2 / N_d$ for most cases, except \textit{outgo-slow} and \textit{ingo}. Comparing to the last test, the amplitude of perturbations matches the original best at $\xi=10$, as not much smoothing is needed with a low resolution in $\lambda$-space. Qualitatively, the output temperature maps match the input fairly well for $\xi \geq 10$. Patterns of \textit{ingo}, \textit{outgo-slow}, and \textit{outgo} are clearly seen for all $\xi$, albeit very smoothed at large $\xi$. In particular, the velocity of the output perturbations at large $\xi$ for \textit{outgo-slow} and \textit{ingo} are larger than that of the input, producing steeper stripes. Thus, while smoothing allows patterns in the maps to been seen more clearly, they also affect certain properties of the perturbations if the effect of the smoothing is significant. As with Fig.~\ref{fig:test_cases}, shapes within the input temperature maps are not resolved well, shown for \textit{two-rings} and \textit{bumps}, due to the lack of resolution in $\lambda$. The success of this method even with data sparsely sampled in frequency-space highlights the results that LSST will produce regarding accretion disk and reverberation mapping campaigns.

\begin{figure*}
    \centering
    \includegraphics[width=\textwidth]{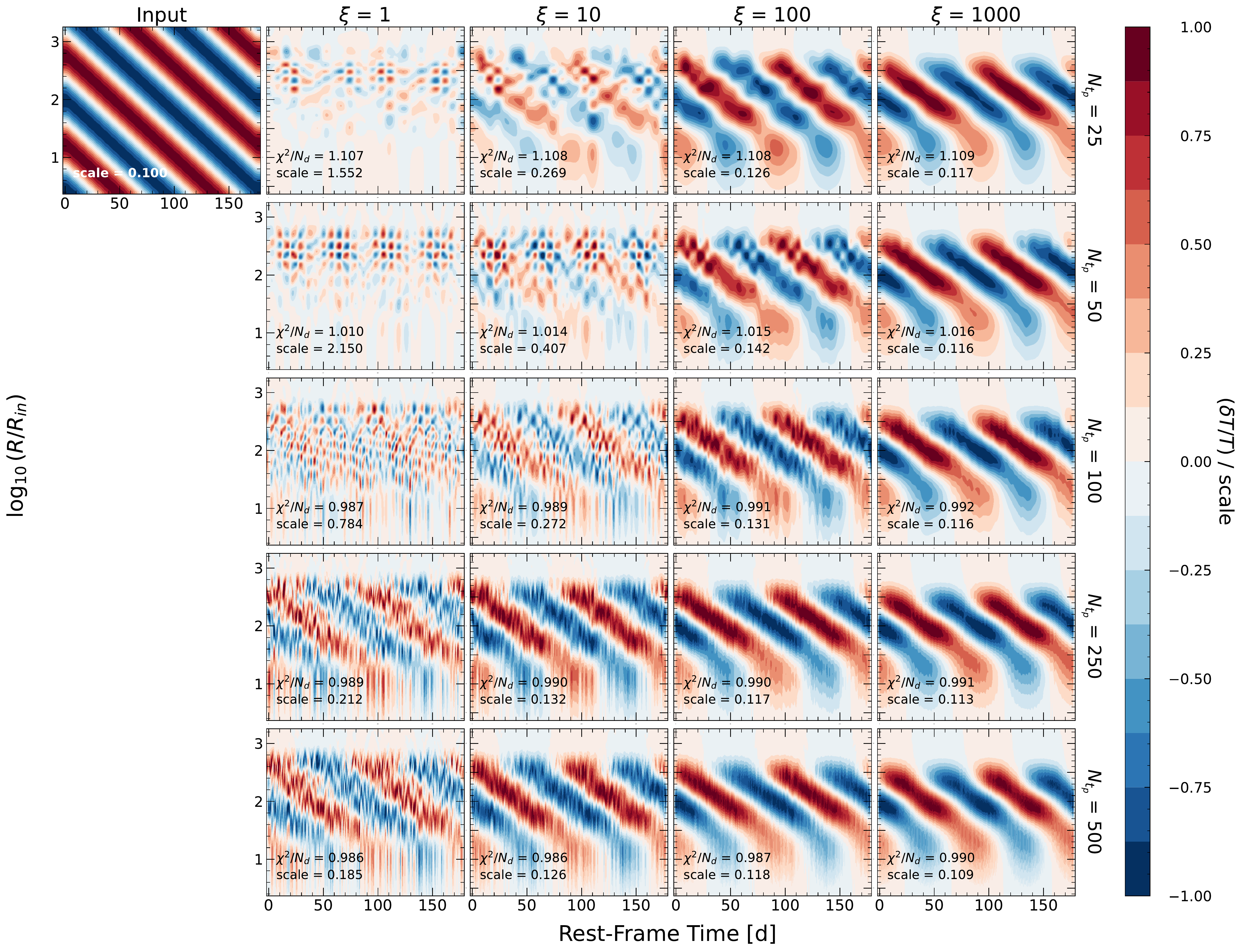}
    \caption{Resolution testing in $t_p$-space for the suite of tests described in Appendix \ref{sec:res_tests}. These tests use the same parameters as those described in Fig.~\ref{fig:test_cases}. These four examples are chosen from a number of tests done to produce the results seen in Fig.~\ref{fig:chi2}. }
    \label{fig:NtpTest}
\end{figure*}

\begin{figure*}
    \centering
    \includegraphics[width=\textwidth]{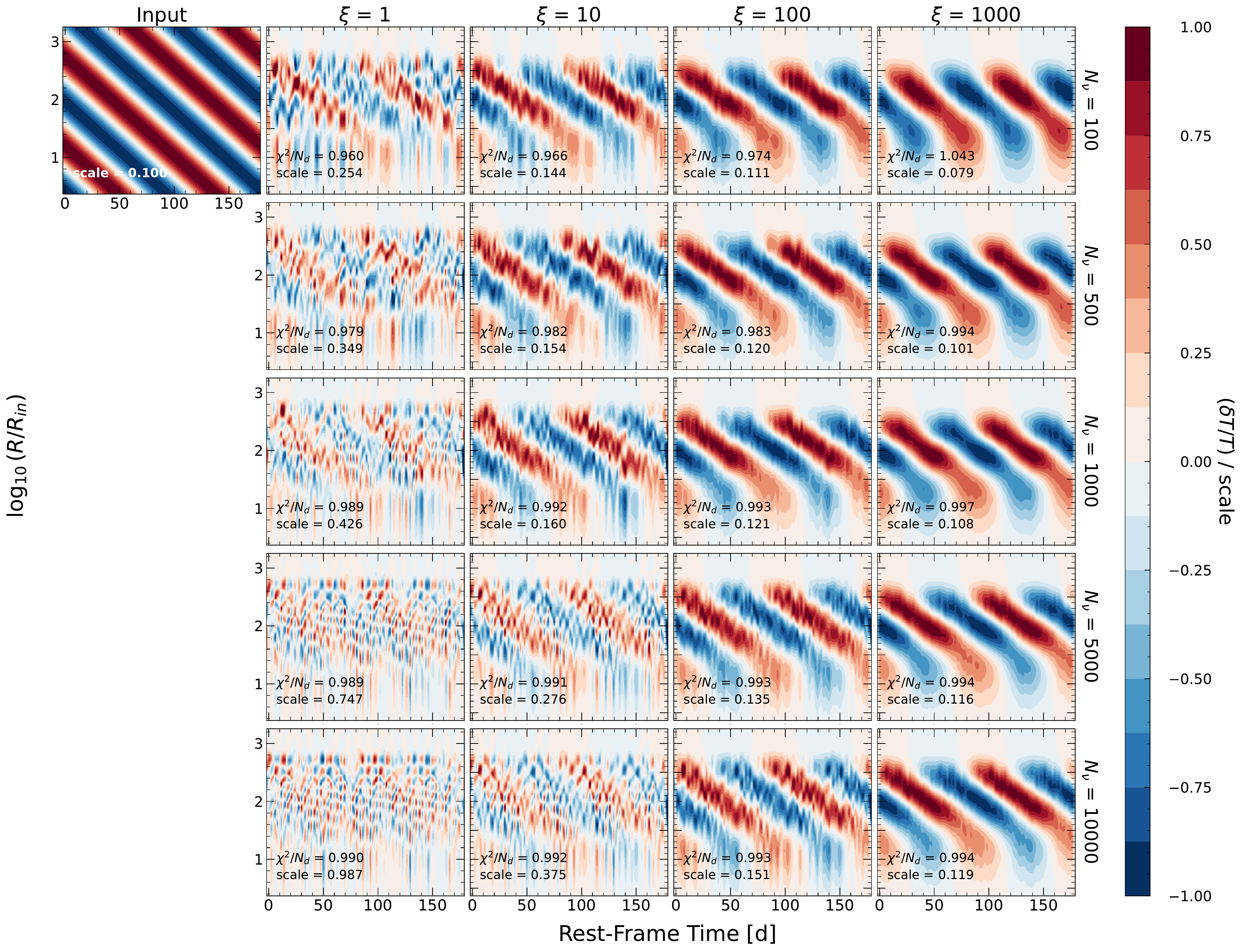}
    \caption{Same as Fig.~\ref{fig:NtpTest}, but testing the spectral resolution ($N_\nu$).}
    \label{fig:NnuTest}
\end{figure*}

\begin{figure*}
    \centering
    \includegraphics[width=\textwidth]{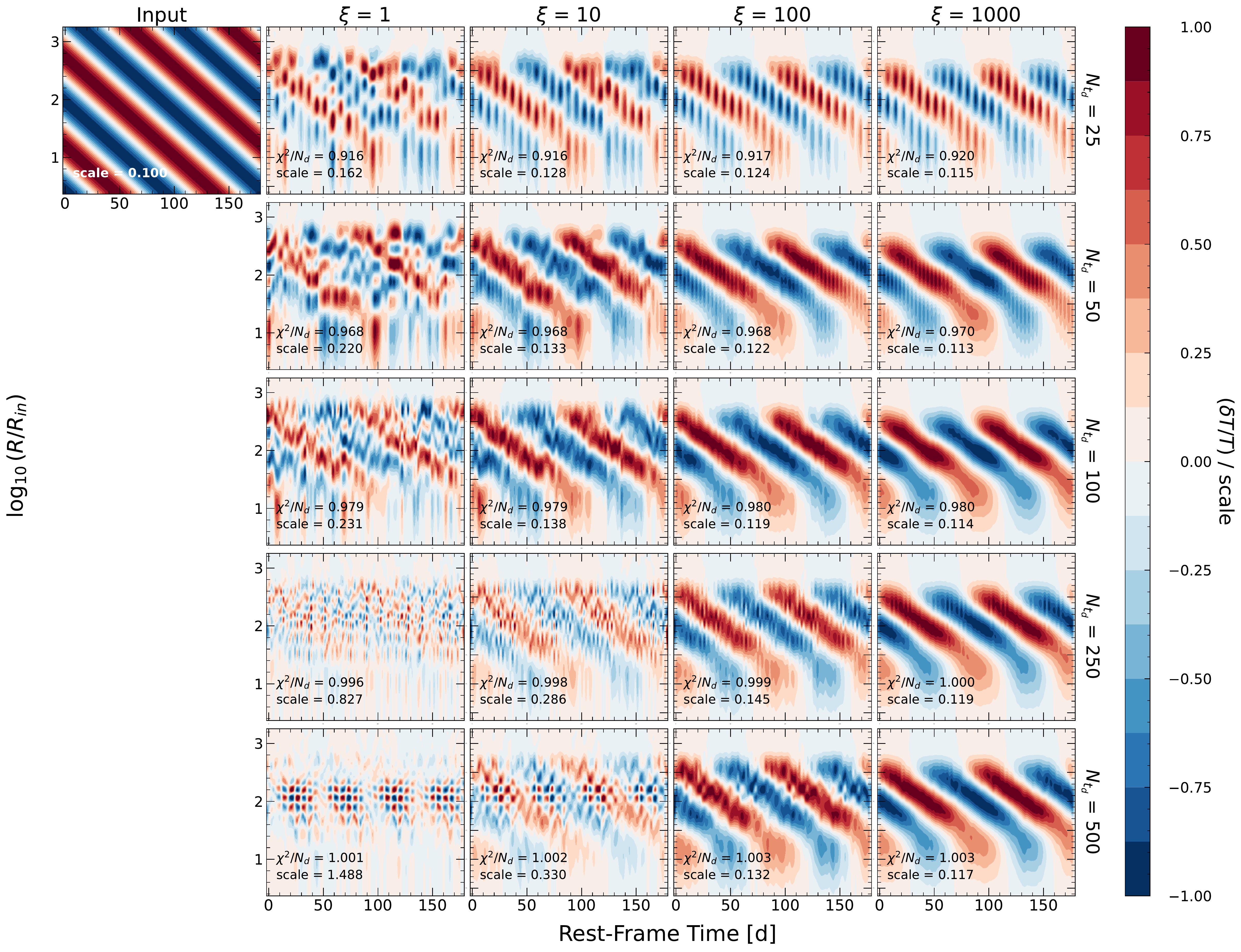}
    \caption{Same as Fig.~\ref{fig:NtpTest}, but testing the cadence ($N_{t_d}$).}
    \label{fig:NtdTest}
\end{figure*}


\begin{figure*}
    \centering
    \includegraphics[width=\textwidth]{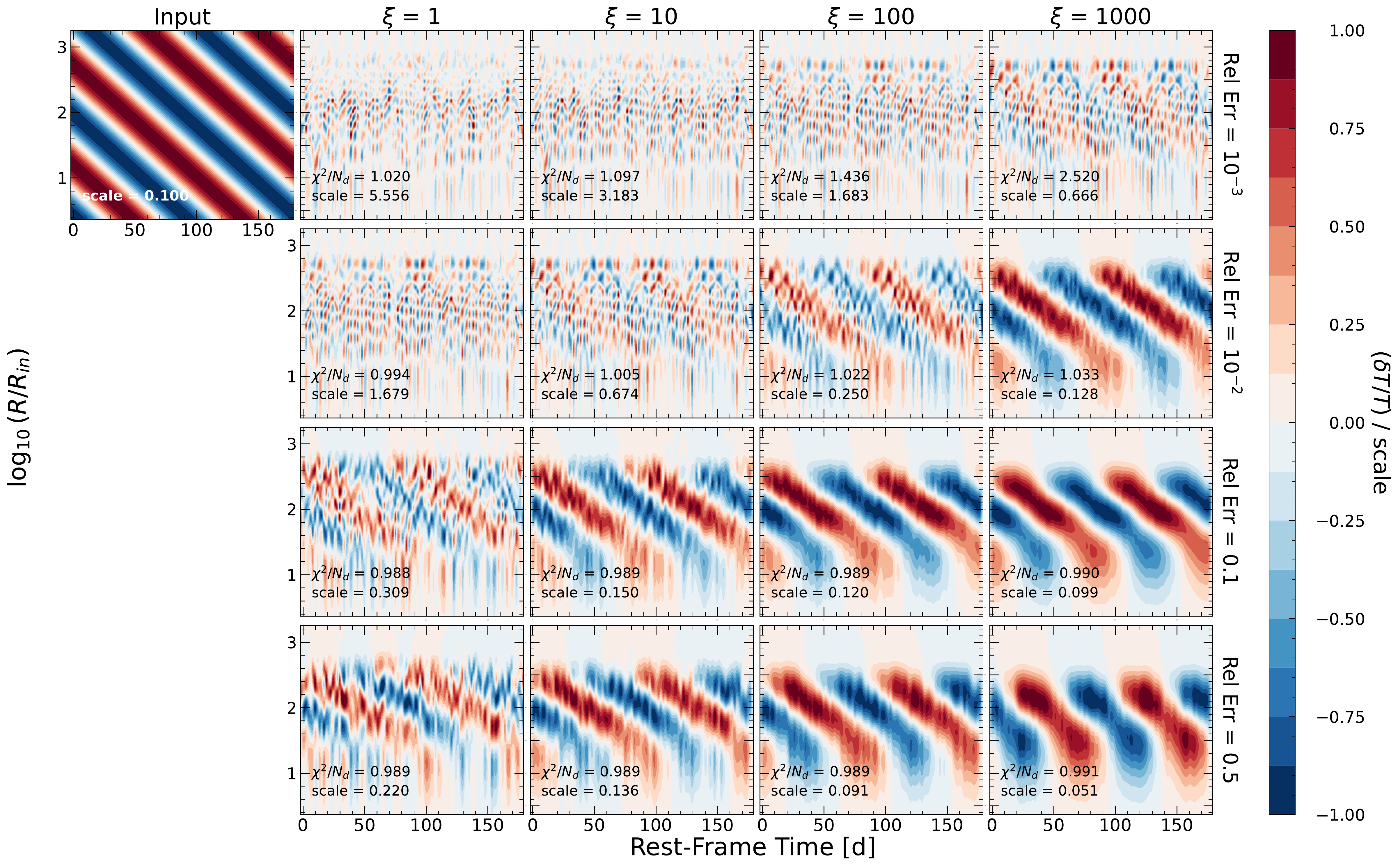}
    \caption{Same as Fig.~\ref{fig:NtpTest}, but testing the effect of the relative error in the input data. In general, the error of the input spectra is relative to the steady-state spectrum $F_{\lambda, SS}$. Each row in the figure represents a test whose error relative to $F_{\lambda, SS}$ is labeled to the right. }
    \label{fig:ErrTest}
\end{figure*}

\section{Testing the Presence of Fast Outgoing Waves}\label{sec:in_and_out_test}

\begin{figure*}
    \centering
    \includegraphics[width=\textwidth]{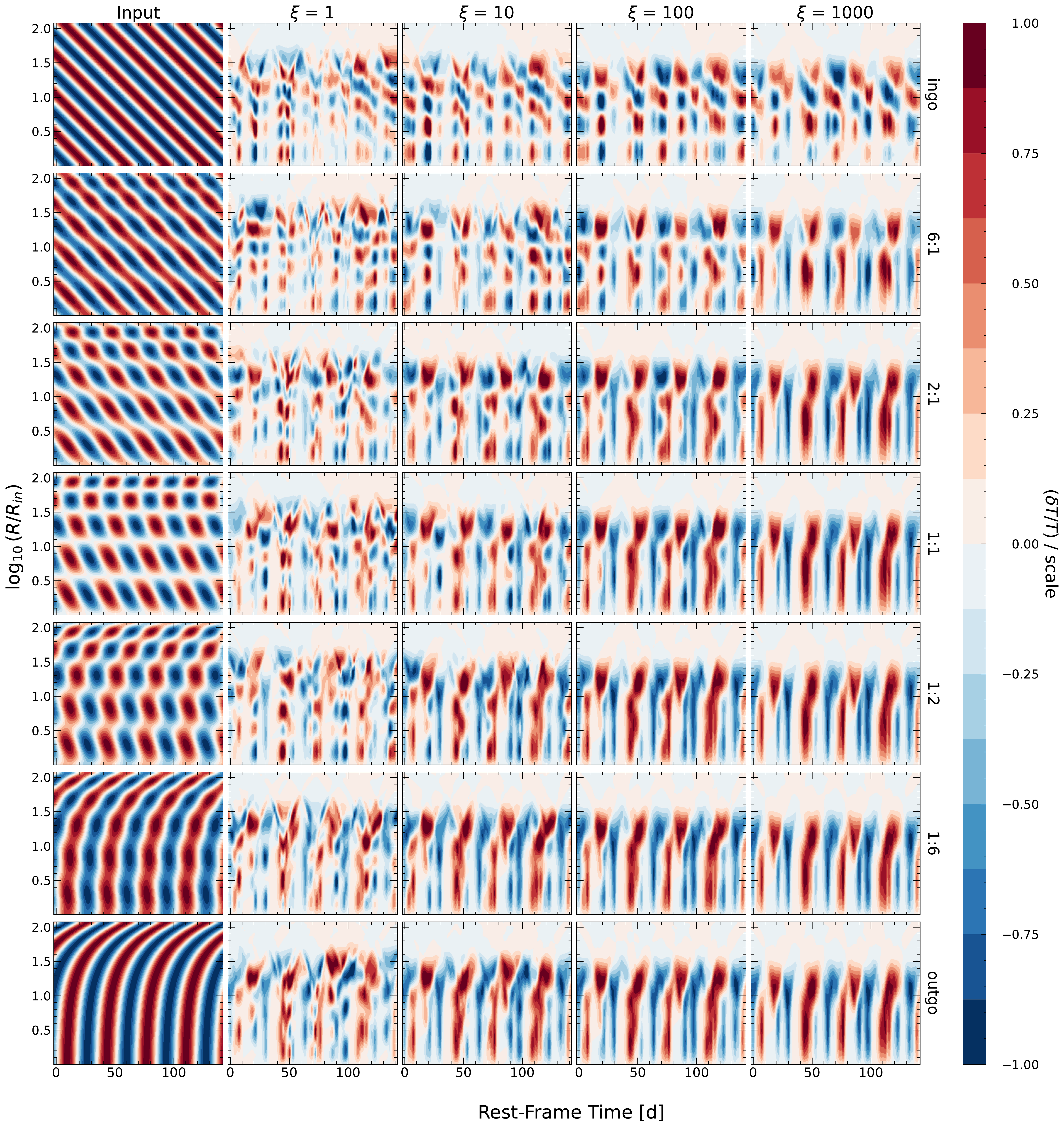}
    \caption{The test described in Appendix \ref{sec:in_and_out_test} to see the visibility of low-amplitude, fast, lamppost-like waves in the SDSS-RM spectra. Each row corresponds to a different ratio of the amplitude of the \textit{ingo} pattern to the \textit{outgo} pattern. The first row is completely \textit{ingo} and the last row is completely \textit{outgo}. Each inversion was performed using the spectral and temporal sampling of RMID085, as well as its AGN parameters.} 
    \label{fig:InAndOutTest}
\end{figure*}

While the results from the SDSS-RM spectra do not reveal any resemblance to fast, outgoing, lamppost-like waves, they may be a superposition of these waves and other perturbations. We perform a test using the same cadence, spectral parameters, and AGN parameters of RMID085 to see if the ingoing/outgoing slow perturbation seen could be a combination of ingoing perturbations and fast, outgoing waves. We perform this test similar to the test done in NK22, with various versions of the \textit{in-and-out} pattern. We input 7 different temperature maps with the \textit{in-and-out} pattern, each with a different ratio of the \textit{ingo} to the \textit{outgo} pattern. 

The results from this test (Fig.~\ref{fig:InAndOutTest}) confirm that the ingoing and outgoing slow perturbations we find in the SDSS-RM data are real and not superimposed on low-amplitude, lamppost-like waves. The test using only \textit{ingo} as the pattern resembles an \textit{ingo} pattern in the output: vertical columns with alternating positive and negative amplitude with patches of the wave on each column. Even when the ratio of \textit{ingo} to \textit{outgo} is 6-to-1, an outgoing wave pattern can still be seen at the highest $\xi$. The highest $\xi$ for all cases, which is not as high as the $\xi$ displaying most patterns in the SDSS-RM data, shows vertical stripes alternating in amplitude as a function of time. Therefore, if there were low-amplitude lamppost-like waves in addition to the large amplitude, slow-moving perturbations, we would see them at large $\xi$ at SDSS-RM cadence and spectral range.

\section{Systematics Due to Light Travel Smearing}\label{sec:smearing}

Here we investigate the systematic bifurcation at large radii, near the edge of the range of radii probed by a given spectral range. We perform three separate tests using the spectral range and AGN parameters of RMID038. We use an input temperature map with the \textit{ingo} pattern and utilize three different samplings: one at the original cadence of RMID038, one sparse sampling with only three observations, and one dense sampling with a cadence of one day. For each test case, we also produce a map of the smearing term developed in \S \ref{sec:methods} (Eqn.~\ref{eqn:smearing_tot}). Note that the light-travel smearing term here is not to be confused with the smoothing term $\xi$ in the matrix inversion. 

\begin{figure*}
    \centering
    \includegraphics[width=.7\textwidth]{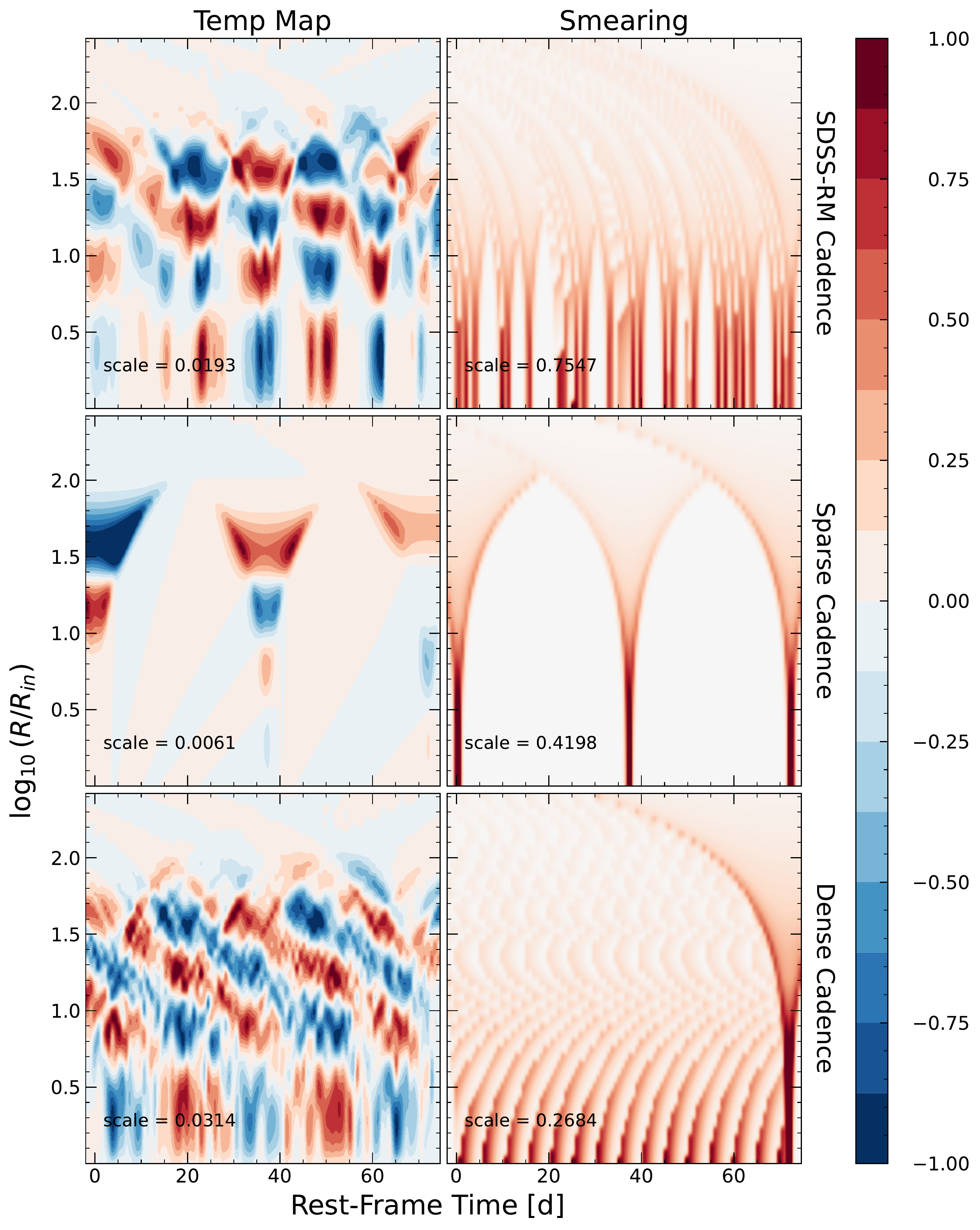}
    \caption{Comparing the systematic ``fanning out" at large radii seen in our output temperature maps to the smearing term from NK22's method. This test was performed using the \textit{ingo} pattern with the cadence, spectral sampling, and AGN parameters of RMID038. Each row of the figure shows a test using different cadences: \textit{top} - same cadence as RMID038, \textit{middle} - only three sampled data points at the two edges of the time range and at the center, \textit{bottom} - the same range as the original sampling with a cadence of 1 day. The left column displays the output temperature maps using $\xi=100$, and the right column shows the smearing terms used to construct the maps. Both the temperature maps and smearing maps are divided by the 99$^{th}$ percentile of the absolute value of the data, labeled as the ``scale" in each panel. }
    \label{fig:smearing_test}
\end{figure*}

The results are shown in Fig.~\ref{fig:smearing_test}, with each map divided by the 99$^{th}$ percentile of the absolute value of the data. The sparse sampling test shows that at each observation $t_d$, a similar pattern emerges with bifurcation at large radii. The shapes within the smearing term map are also seen in the exact same manner in the output temperature map. As the cadence decreases for the original RM sampling, the ``fanning out" pattern for multiple observations begin to overlap at large radii. This allows the map to blend these systematic patterns, averaging them out over a range of $t_p$. As the resolution in $t_d$ grows for the densely sampled case, smearing term patterns for individual observations are nearly impossible to distinguish. These smearing terms combine at all radii and average out over the entire map. The resulting temperature map also shows very little remnants of the smearing term at this cadence.

The significance of the smearing term also changes as the number of observations grows. The scale of the smearing is greatest when the temporal sampling is somewhat low, similar to the cadence of the original RM spectra. As the cadence increases, the smearing term is averaged out at all radii, causing the scale of the smearing term to decrease. Though, in general the significance and impact of this artifact of the smearing term depends on the parameters of the AGN. For many SDSS-RM quasars, a width of one fan can span the entire baseline of observations. Inspecting the smearing term in Eqn.~(\ref{eqn:smearing_tot}), this depends on the range of probed radii, the binning in $t_p$-space, and the BH mass of the AGN.

To further demonstrate that these bifurcations at large radii are not real, but systematic, we perform a ``garbage-in, garbage-out" test. We utilize spectra from one of the objects in our SDSS-RM sample, but randomly ``scramble" the order of the spectra. We scramble the spectra several times and obtain a resulting temperature map for each set of spectra. If these features were real, we would not see them in the scrambled temperature maps, but only in the original map. However, we find that the bifurcations are seen in all temperature maps from scrambled spectra. In addition, the scale for the scrambled temperature maps are unphysical, reaching $\sim 1.5T_0$, and the $\chi^2 / N_d$ are much larger than that for the original spectra.

\end{appendix}


\bsp	
\label{lastpage}
\end{document}